\documentclass[10pt,letterpaper,twocolumn,aps,manuscript,prb,superscriptaddress,twocolumns,showpacs]{revtex4}
\usepackage[T1]{fontenc}
\usepackage[utf8]{inputenc}
\usepackage{color}
\usepackage{bm}
\usepackage{amstext}
\usepackage{amssymb}
\usepackage{graphicx}
\usepackage{amsmath}
\usepackage {hyperref}

\DeclareUnicodeCharacter{00A0}{~}

\makeatletter

\pdfpageheight\paperheight
\pdfpagewidth\paperwidth

\providecommand{\tabularnewline}{\\}

\@ifundefined{textcolor}{}
{%
 \definecolor{BLACK}{gray}{0}
 \definecolor{WHITE}{gray}{1}
 \definecolor{RED}{rgb}{1,0,0}
 \definecolor{GREEN}{rgb}{0,1,0}
 \definecolor{BLUE}{rgb}{0,0,1}
 \definecolor{CYAN}{cmyk}{1,0,0,0}
 \definecolor{MAGENTA}{cmyk}{0,1,0,0}
 \definecolor{YELLOW}{cmyk}{0,0,1,0}
}

\PassOptionsToPackage{caption=false}{subfig}

\newcommand{\ket}[1]{\left| #1 \right>}
\newcommand{\bra}[1]{\left< #1 \right|}
\newcommand{\inner}[2]{\left< #1 \big|#2\right>}
\newcommand{\expect}[1]{\left< #1 \right>}

\newcommand{\up}[0]{\ket{\uparrow}}
\newcommand{\down}[0]{\ket{\downarrow}}

\let\oldsqrt\sqrt
\def\sqrt{\mathpalette\DHLhksqrt}
\def\DHLhksqrt#1#2{%
\setbox0=\hbox{$#1\oldsqrt{#2\,}$}\dimen0=\ht0
\advance\dimen0-0.2\ht0
\setbox2=\hbox{\vrule height\ht0 depth -\dimen0}%
{\box0\lower0.4pt\box2}}

\@ifundefined{showcaptionsetup}{}{%
 \PassOptionsToPackage{caption=false}{subfig}}
\usepackage{subfig}

\makeatother

\begin{document}

\title{Exchange-based two-qubit gate for singlet-triplet qubits}

\author{Matthew P. Wardrop}

\author{Andrew C. Doherty}

\affiliation{Centre for Engineered Quantum Systems, School of Physics, The University of Sydney, Sydney, NSW 2006, Australia}


\date{\today}

\begin{abstract} We analyse a simple exchange-based two-qubit gate
for singlet-triplet qubits in gate-defined semiconductor quantum dots that can
be implemented in a single exchange pulse. Excitations from the logical subspace
are suppressed by a magnetic field gradient that causes spin-flip transitions to
be non-energy-conserving. We show that the use of adiabatic pulses greatly
reduces leakage processes compared to square pulses. We also characterise the
effect of charge noise on the entanglement fidelity of the gate both
analytically and in simulations; demonstrating high entanglement fidelities for
physically realistic experimental parameters. Specifically we find that it is
possible to achieve fidelities and gate times that are comparable to single-
qubit states using realistic magnetic field gradients. \end{abstract} \maketitle

\section{Introduction}

Semiconductor quantum dot systems have become an increasingly promising
architecture for large-scale quantum computing \cite{Hanson2007,Kloeffel2013},
growing out of the seminal work of Loss and DiVincenzo \cite{Loss1998}.
Any successful quantum computing architecture must, with high reliability
and precision, be able to: encode information, perform universal logical
operations, generate measurable results, and be scalable to allow
for large computations \cite{DiVincenzo2000}. While various semiconductor
materials have yielded promising results, including among others silicon
\cite{Maune2012,Dehollain2014} and carbon \cite{Trauzettel2007,Bulaev2008}
based structures; GaAs/AlGaAs heterostructures remain very popular
due to the advanced techniques developed for this material by experimenters.

The original semiconductor proposal \cite{Loss1998} recognised the
two-level spin system of an electron localised in a gate-defined semiconductor
quantum dot as a natural encoding of a qubit, which is now called
the Loss-DiVincenzo qubit. Two-qubit control was to be provided by
exchange coupling between the dots, which has been implemented by
modifying the gate voltages that define the dots \cite{Petta2005};
and is now a matter of routine practice. Single-qubit control is more
challenging \cite{Koppens2006,Pioro-Ladriere2008,Obata2010,Nadj-Perge2010,Brunner2011},
and is usually implemented using electrically driven spin resonance
in the presence of magnetic field gradients that allow for individual
addressing.

Various modifications to Loss-DiVincenzo qubits and their manipulation
have been proposed, each trading off the relative simplicity of single
electron spin qubit encoding for systems of greater redundancy, ease
of implementation, and/or resilience to experimental noise. Among
the most promising of these new proposals are the ``singlet-triplet''
qubits \cite{Levy2002,Johnson2005,Hanson2007a,Barthel2009,Bluhm2010a,Barthel2012,Shulman2012,Stepanenko2012,Klinovaja2012,Dial2013},
which will be the focus of this paper. Another promising candidate
is the exchange-only qubit \cite{DiVincenzo2000a,Laird2010,Gaudreau2011,Medford2013,Medford2013a,Doherty2013},
which encodes logical qubits in the spins of three electrons; allowing
for full electronic control through the exchange interaction alone.

Singlet-triplet qubits encode logical information in a pair of electron
spins. The logical subspace of these qubits is the two-dimensional
subspace of a pair of electron spins that is not Zeeman-shifted in
an applied magnetic field, making them resistant to global magnetic
field fluctuations \cite{Levy2002}. Single qubit
operations are performed using a (potentially static) magnetic field
gradient and an exchange coupling between the dots. We describe these
qubits in more detail in section \ref{sub:Single-Qubit-Gates}. Static
magnetic field gradients have been demonstrated using dynamic nuclear
polarisation\cite{Foletti2009,Gullans2010} and patterned nano-magnets\cite{McNeil2010,Takakura2010};
with gradients as large as $100\, mT$. There have been several proposals
for two-qubit operations, the realisation of any being sufficient
for universality of quantum computation \cite{Lloyd1995}. The only
two-qubit gate currently demonstrated in experiment uses capacitive
coupling \cite{Weperen2011,Shulman2012}.

In this paper we present a proposal for an exchange-based two-qubit
gate for neighbouring singlet-triplet qubits that can effect high
fidelity operations in a single adiabatic pulse. The use of exchange
coupling has the significant advantage that gates can be fast, with
gate times comparable to single qubit operations. However, use of
exchange coupling between singlet-triplet qubits typically causes
spin-flip transitions that result in excursions from the qubit subspace,
leading to so-called leakage errors. Such leakage errors are suppressed
during our gate by a static magnetic field gradient that causes spin-flip
transitions to violate energy conservation; and are further mitigated
by the adiabatic pulsing of the inter-qubit exchange couplings. Our
proposal does not depend on the details of the substrate in which
the quantum dots are embedded, or the way in which the exchange coupling
and magnetic field gradients are realised; allowing for novel effective
fields and couplings to be used (e.g. \cite{Coello2010,Mehl2014}).
In simulations incorporating physically realistic charge noise, we
found that with static magnetic field gradients less than $100\, mT$,
and gate times as short as $7\, ns$, our gate can perform with entanglement
fidelities in excess of 99.9\%. In this regime, our gate performs
with similar fidelity to single qubit operations. Our study complements
a similar proposal described by Klinovaja and collaborators \cite{Klinovaja2012},
which considers pulse sequences as an alternative to adiabatic pulses for solving the
problems of leakage, and focusses on spin orbit coupling and Overhauser
noise instead of charge noise. In addition, Li and collaborators \cite{Li2012} describe several
pulse sequences which can effect two-qubit
gates; and Kestner \cite{Kestner2013}, Wang \cite{Wang2014} and collaborators have
developed pulse sequences which mitigate the effect of low-frequency Overhauser and charge noise. The notion of energetically suppressing
leakage processes also appears in our two-qubit gate proposal \cite{Doherty2013}
for the resonant exchange qubit \cite{Taylor2013,Medford2013}, in
which context the use of adiabatic pulses is also expected to lead to
significant reduction in leakage.

Our two-qubit gate does not have some of the drawbacks of earlier
proposals. The exchange-based two-qubit gate accompanying the original
singlet-triplet qubit proposal \cite{Levy2002} required a sequence
of complicated exchange pulses between neighbouring singlet-triplet
qubits. Apart from their complexity, these sequences also required
very precise timing and negligible charge noise in order to minimise
leakage. Capacitive two-qubit gate proposals \cite{Taylor2005,Weperen2011,Shulman2012,Trifunovic2012}
have the advantage of having actually been implemented, and being
applicable for more widely spaced qubits; but it has proven difficult
thus far to create large charge dipoles in singlet-triplet qubits,
leading to gate times an order of magnitude slower than single qubit
exchange gates \cite{Petta2005,Weperen2011}. There have been some
promising proposals to strengthen capacitive interactions; such as
floating gates \cite{Trifunovic2012}. A more fundamental
limitation is that charge noise in the control voltages couples into
capacitive interactions unfavourably \cite{Shulman2012}. In contrast,
our proposal promises gates that can be implemented between nearest
neighbours using relatively simple adiabatic pulses; with gate times
comparable to single qubit operations, and a more favourable noise
scaling that allows one to trade off gate speed for less sensitivity
to charge noise.

This paper is organised as follows: in section \ref{sec:Model} we
describe a model for semiconductor quantum dot systems; in section
\ref{sec:Gate-Operations} we describe the mechanics of our two-qubit
operation; in section \ref{sec:Sources-of-Error} we introduce a model
for charge noise in singlet-triplet systems; in section \ref{sec:Quantifying-Gate-Performance}
we analytically investigate the performance of our gate subject to
this charge noise model; in section \ref{sec:Simulations} we present
the results of simulations of our two-qubit gate and compare with
the results of the previous section; and in section \ref{sec:Discussion}
we discuss the significance of these results.

\section{Physical Model\label{sec:Model}}

\begin{figure}
\includegraphics[width=1\columnwidth]{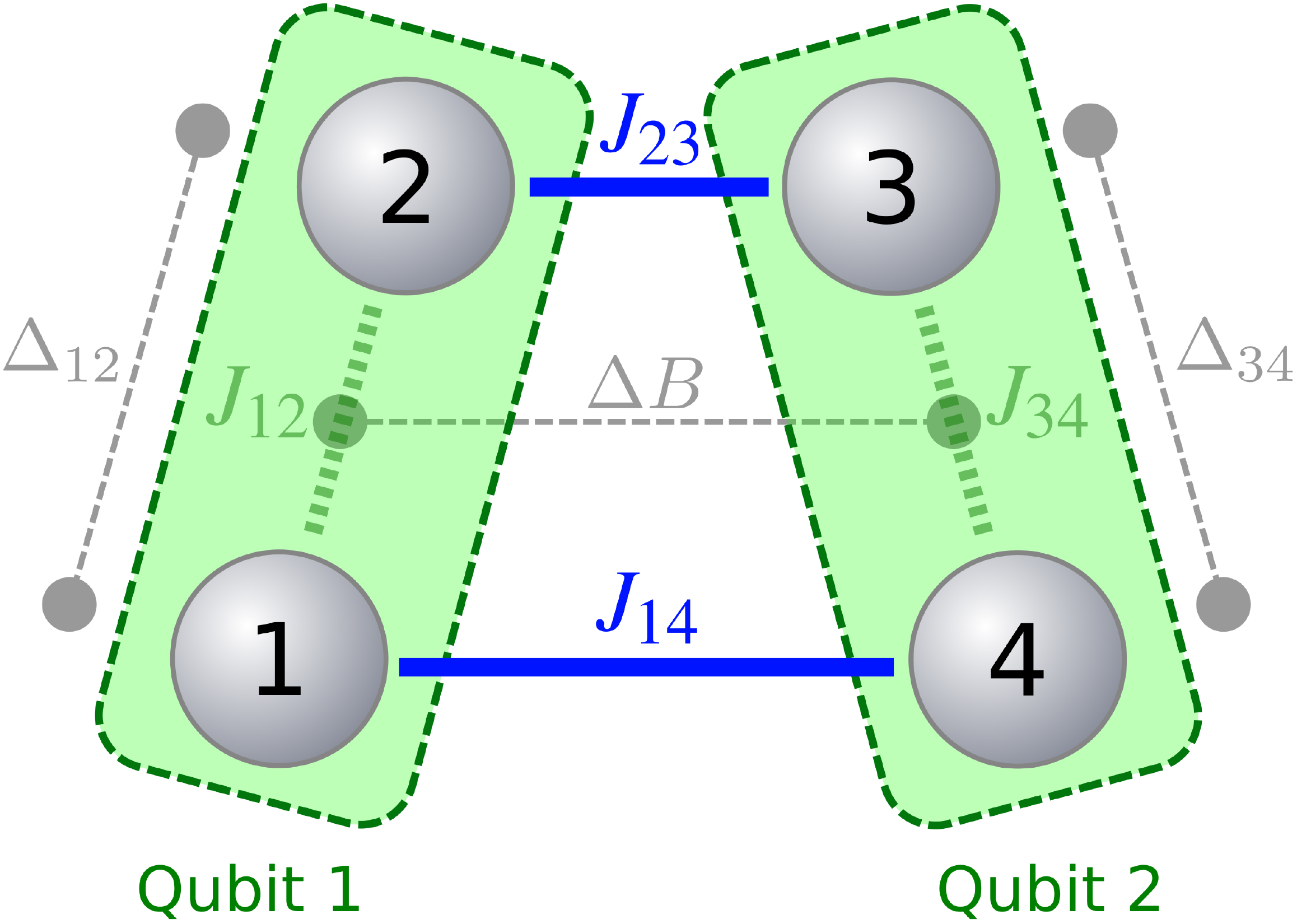}

\protect\caption{(colour online) Schematic diagram of a two singlet-triplet qubit configuration.
 The four quantum dots are indexed by $\{1,2,3,4\}$; and are slanted
to indicate that a specific physical arrangement is unimportant. The
intra-qubit exchange couplings required for single qubit operations
are shown in dashed green ($J_{12}$ and $J_{34}$); and the inter-qubit
couplings required for our two-qubit gate operation are shown in solid
blue ($J_{14}$ and $J_{23}$). Important magnetic field gradients
are depicted by thin broken grey lines joining solid discs; and labelled
with the appropriate symbols: $\Delta B=\frac{1}{2}\left(B_{1}+B_{2}-B_{3}-B_{4}\right)$
and $\Delta_{ij}=B_{i}-B_{j}$.\label{fig:Square-arrangement}}
\end{figure}

In this section we introduce the model we use to describe exchange-coupled
quantum dots; which we use in the following section to explain the
mechanism of our proposed two qubit gate for singlet-triplet qubits.

We have chosen to model a system of $N$ electrons (each isolated
in a gate-defined quantum dot) with the time dependent Heisenberg
Hamiltonian:
\begin{equation}
H(t)=\mu\sum_{n=1}^{N}B_{n}(t)\sigma_{z}^{n}+\frac{1}{4}\sum_{\left<i,j\right>}J_{ij}(t)(\bm{\sigma}^{i}\cdot\bm{\sigma}^{j}-I),\label{eq:simple_hamiltonian}
\end{equation}
where $\left<i,j\right>$ indicates that the sum should only include
pairs of $i$ and $j$ if there exists non-negligible quantum tunnelling
between quantum dots $i$ and $j$; and $\bm{\sigma}^{i}$ is the
vector of Pauli operators $(\sigma_{x},\sigma_{y},\sigma_{z})$ acting
on dot $i$. The first sum of terms describes Zeeman splitting of
the spin states at each dot due to the local magnetic field $B_{n}(t)$;
and the second describes the exchange couplings $J_{ij}(t)$ between
the dots. We set $\hbar=1$ throughout this paper.

A large external magnetic field $B_{0}=\bm{B}_{0}^{z}=\frac{1}{4}\sum_{n}B_{n}$
creates a preferred orientation, which we arbitrarily label the $z$-axis.
All of the magnetic fields $B_{n}$ are taken to be along this $z$-axis,
as the effects of perpendicular fields will be suppressed provided
$\bm{B}_{n}^{\perp}\ll\bm{B}_{n}^{z}$. We will find it useful to
consider the magnetic fields in the following basis: the global background magnetic
field $B_{0}=\frac{1}{4}\sum_{n}B_{n}$, the inter-qubit gradient
$\Delta B=\frac{1}{2}\left(B_{1}+B_{2}-B_{3}-B_{4}\right)$, and the
intra-qubit gradients $\Delta_{12}$ and $\Delta_{34}$ with $\Delta_{ij}=B_{i}-B_{j}$.
Thus $B_{1}=B_{0}+\frac{1}{2}\Delta B+\frac{1}{2}\Delta_{12}$; and
so on.

Computation using singlet-triplet qubits requires control of the exchange
couplings $J_{ij}(t)$, which depend on the shape of the quantum dot
potential wells that are in turn determined by electrode voltages
that we parameterise by $\varepsilon_{ij}$; and so $J_{ij}(t)=J_{ij}(\varepsilon_{ij}(t))$.
The precise dependence of $J_{ij}$ on $\varepsilon_{ij}$ is determined
by the microscopic details of experimental apparatus. In order to
make quantitative statements about our proposal, in section \ref{sub:Charge-Noise}
we will consider a phenomenological fit to data from GaAs/AlGaAs singlet-triplet
experiments.

This model can be regarded as an approximation of the more general
Hubbard model with $N$ sites, local magnetic fields $\bm{B}_{n}(t)$
and tunnelling between sites $i$ and $j$ of $t_{ij}(t)$. The exchange
coupling terms $J_{ij}(\bm{\sigma}\cdot\bm{\sigma}-II)$ are the second
order perturbative effect of quantum tunnelling $t_{ij}$; with $J_{ij}=4t_{ij}^{2}/E_{C}$,
where $E_{C}$ is the energy penalty associated with charging a quantum
dot with two electrons. This approximation holds in the limit of weak
tunnelling $t_{ij}\ll E_{\textrm{C}}$.

\section{Logical Operations\label{sec:Gate-Operations}}

In this section we provide an intuition for how our gate works; before
describing it in detail. The key physics that underpins the operation
of our gate is the same as for single-qubit exchange gates; made more
complicated by the possibility of low-energy excitations from the
logical subspace. We suppress these by applying a gradient magnetic
field to make spin-flip transitions non-energy-conserving; as also
discussed in Klinovaja et al. \cite{Klinovaja2012}. We first review
single qubit gates.

\subsection{Single Qubit Gates for Singlet-Triplet Qubits\label{sub:Single-Qubit-Gates}}

Singlet-triplet qubits are encoded in the spins of two electrons,
each isolated in a quantum dot (such as one of the qubits in figure
\ref{fig:Square-arrangement}); and are controlled using inter-dot
magnetic field gradients and variable exchange coupling, in the presence
of a strong global magnetic field. The Hamiltonian describing such
a system is that given by equation (\ref{eq:simple_hamiltonian})
restricted to two dots ($N=2$).

The strong global magnetic field $B_{0}$ makes the spin basis a natural
one for this system: $\ket{\uparrow\uparrow}$, $\ket{\uparrow\downarrow}$,
$\ket{\downarrow\uparrow}$, and $\ket{\downarrow\downarrow}$, where
the arrows indicate the $\hat{S}^{z}$ projection of the electrons'
spin. The global field $B_{0}$ Zeeman splits the $\sum\hat{S}^{z}\ne0$
states from the $\sum\hat{S}^{z}=0$ states, which energetically suppresses
the hyperfine interactions between the electron and semiconductor
lattice nuclear spins that would otherwise cause excitations between
them \cite{Petta2005}. The singlet-triplet qubit is encoded in the
two-dimensional $\sum\hat{S}^{z}=0$ subspace, which is spanned by
the states $\{\ket{\uparrow\downarrow},\ket{\downarrow\uparrow}\}$.
The exchange term $\frac{1}{4}J_{12}(\bm{\sigma}^{1}\cdot\bm{\sigma}^{2}-I)$
has two eigenstates in the logical subspace: the singlet state $\ket{S}=\left(\ket{\uparrow\downarrow}-\ket{\downarrow\uparrow}\right)/\sqrt{2}$
and the $\hat{S}_{z}=0$ triplet state $\ket{T_{0}}=\left(\ket{\uparrow\downarrow}+\ket{\downarrow\uparrow}\right)/\sqrt{2}$;
which are customarily chosen to be the computational basis states
(hence the name ``singlet-triplet'' qubit\cite{Petta2005}).

\begin{figure}
\includegraphics[width=0.75\columnwidth]{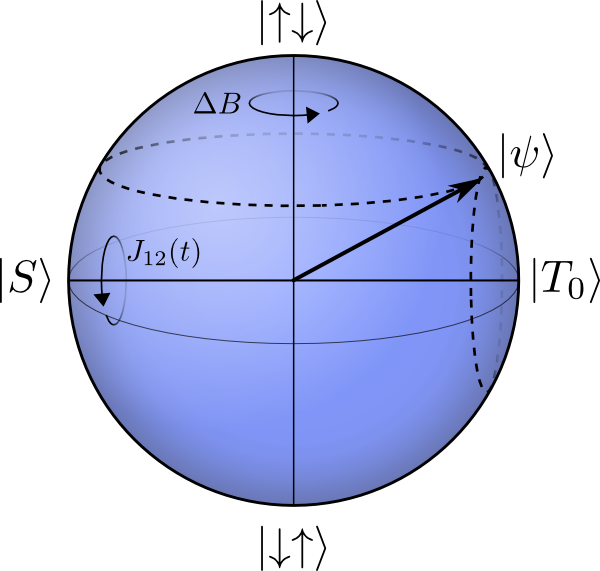}

\protect\caption{(colour online) Schematic showing the alignment of the single-qubit Bloch sphere. The magnetic field
gradient $\Delta B$ rotates an arbitrary qubit state $\ket{\psi}$ about the z-axis of the Bloch sphere, while the exchange interation
$J_{12}$ rotates it about the x-axis.\label{fig:dqd_bloch}}
\end{figure}

Universal control of a qubit entails the ability to perform arbitrary rotations
of Bloch vectors around the Bloch sphere, which requires two independent axes of
rotation. For singlet-triplet qubits, these are provided by a static magnetic
field gradient $\Delta B = B_1 - B_2$ and a variable exchange coupling
$J_{12}(t)$, as depicted in figure \ref{fig:dqd_bloch}. In order to simplify discussion in this paper, we have chosen to
orient the Bloch sphere such that the north and south poles are aligned with
$\ket{\uparrow\downarrow}$ and $\ket{\downarrow\uparrow}$ respectively; and the
x-axis with the singlet state
$\ket{S}=\left(\ket{\uparrow\downarrow}-\ket{\downarrow\uparrow}\right)/\sqrt{2}$
and the triplet state
$\ket{T_{0}}=\left(\ket{\uparrow\downarrow}+\ket{\downarrow\uparrow}\right)/\sqrt{2}$.
This choice of Bloch axes is unconventional (e.g. \cite{Petta2005}); but allows
us to describe the operation of our two-qubit gate in terms of diagonal Pauli
z-operators in subsequent sections. In this basis, the magnetic field gradient
causes coherent phase evolution of an arbitrary superposition $\ket{\psi} =
\alpha \ket{\uparrow\downarrow} + \beta \ket{\downarrow\uparrow} \mapsto \alpha
\ket{\uparrow\downarrow} + \beta \exp(i \mu\Delta B t)\ket{\downarrow\uparrow}$,
which describes rotations about the z-axis of the Bloch sphere. It is sufficient
for this field gradient to be static, since one can keep track of the
precession. The exchange operator lowers the singlet state $\ket{S}$ in energy
by a controllable amount $J_{12}(t)$ compared to the triplet states
$\{\ket{\uparrow\uparrow},\ket{T_{0}},\ket{\downarrow\downarrow}\}$. This causes
coherent phase evolution of an arbitrary superposition $\ket{\psi} = \alpha
\ket{S} + \beta \ket{T_0} \mapsto \alpha \ket{S} + \beta \exp(-i \int_0^t
J_{12}(t^\prime) dt^\prime)\ket{T_0}$, which describes x-axis rotations around
the Bloch sphere. When $\ket{\psi}$ is an equal superposition of $\ket{S}$ and
$\ket{T_0}$, these rotations are manifest as coherent oscillations between the
$\ket{\uparrow\downarrow}=\ket{T_0}+\ket{S}$ and
$\ket{\downarrow\uparrow}=\ket{T_0}-\ket{S}$ states. Together, these two
operations can effect an arbitrary rotation in the Bloch sphere, and thus
provide universal control.

To characterise single-qubit gate times for later comparison to two-qubit
operations, we consider an application of a SWAP gate that has the effect of flipping
the spins of the two electrons encoding the qubit state. This occurs
when the qubit state $\ket{\psi}$ is rotated about the x-axis of the Bloch sphere
by $\pi$ radians, which is when $\int_{0}^{\tau}J_{12}(t)dt=\pi$ (c.f. \cite{Petta2005}),
or when gate time $\tau=\pi/J_{12\,\mathrm{avg}}$.

\subsection{Our Two Qubit Exchange Gate\label{sub:Our-Two-Qubit}}

The premise of our two-qubit gate proposal is to use exchange couplings
$J_{14}$ and $J_{23}$ between two singlet-triplet qubits (as shown
in figure \ref{fig:Square-arrangement}) in order to perform a conditional
phase gate (CPHASE) between the logical states of the two qubits.

There are four quantum dots in the combined two-qubit system, which
are described by the Hamiltonian in equation \ref{eq:simple_hamiltonian}
with $N=4$. Just as for the single qubit case, the strong background
magnetic field makes the $2^{4}=16$ spin configurations a natural
basis. The logical basis is the tensor product of the single qubit
subspaces: $\{\ket{\uparrow\downarrow,\uparrow\downarrow},\ket{\uparrow\downarrow,\downarrow\uparrow},\ket{\downarrow\uparrow,\uparrow\downarrow},\ket{\downarrow\uparrow,\downarrow\uparrow}\}$.
Unlike the single qubit case, the logical subspace is not energetically
isolated by the global magnetic field. There are two non-logical states
(called ``leakage'' states) which also have $\sum\hat{S}^{z}=0$
: $\ket{\uparrow\uparrow,\downarrow\downarrow}$ and $\ket{\downarrow\downarrow,\uparrow\uparrow}$.
This makes the system susceptible to zero-energy excitations from
the logical subspace. To make matters worse, such leakage transitions
are actually driven from the logical states $\ket{\uparrow\downarrow,\uparrow\downarrow}$
and $\ket{\downarrow\uparrow,\downarrow\uparrow}$ by the exchange
couplings $J_{14}$ and $J_{23}$ that are necessary for our two-qubit
gate.

The addition of a magnetic field gradient $\Delta B$ between the
two qubits isolates the logical subspace from the leakage states by
approximately $\mu\Delta B$. In the limit that exchange couplings
$J_{14}$ and $J_{23}$ are much less than $\mu\Delta B$, transitions
from the logical subspace to the unwanted $\sum\hat{S}^{z}=0$ states
are energetically forbidden (and thus suppressed). We will later discuss
the use of adiabatic activation of $J_{ij}$ to further suppress leakage.
Note that the field gradient must be much smaller than the applied
homogeneous field $\Delta B\ll B_{0}$, so that $B_{0}$ remains the
dominant energy scale. Since leakage from the logical subspace can
in principle be made negligible by choosing small enough $J_{ij}\ll\mu\Delta B\ll\mu B_{0}$,
we postpone further discussion of leakage until section \ref{sub:Leakage};
and focus on perfectly adiabatic gate operation.

During the operation of our gate, we turn off intra-qubit exchange
couplings $J_{12}$ and $J_{34}$; before activating $J_{14}$ and/or
$J_{23}$. This effectively decouples the system into a new pairing
of quantum dots which are described by exactly the same Hamiltonian
as singlet-triplet qubits, but which are not confined to the singlet-triplet
logical subspace. In particular, notice that if the two qubit system
is initially in the logical state $\ket{\uparrow\downarrow,\downarrow\uparrow}$,
then the new pairings would lead to two-quantum-dot triplet states
$\ket{\uparrow\uparrow}$ and $\ket{\downarrow\downarrow}$ that previously
did not correspond to logical states.

As described in section \ref{sub:Single-Qubit-Gates}, activating
exchange coupling between dots $i$ and $j$ lowers the singlet energy
state by $J_{ij}$ compared to the relevant triplet states. Under
$J_{14}$ and/or $J_{23}$, the logical states that have singlet character
under the new pairings, $\ket{\uparrow\downarrow,\uparrow\downarrow}$
and $\ket{\downarrow\uparrow,\downarrow\uparrow}$, reduce in energy
by approximately $\frac{1}{2}(J_{14}+J_{23})$ compared to the other
two logical states, $\ket{\uparrow\downarrow,\downarrow\uparrow}$
and $\ket{\downarrow\uparrow,\uparrow\downarrow}$. This interaction
looks like a logical $\sigma_{z}\sigma_{z}$ coupling between the
two singlet-triplet qubits; which is well known to generate CPHASE
gates modulo single qubit z-rotations (e.g. \cite{DiCarlo2009}) when
the phase associated with $\sigma_{z}\sigma_{z}$ has accumulated
to $\pi/2$. We will show more rigorously in the following that a
CPHASE gate results after a time $\tau$ such that $\int_{0}^{\tau}\left[J_{14}(t)+J_{23}(t)\right]dt=\pi$;
or $\tau=\pi/(J_{14}+J_{23})_{\textrm{avg}}$. Although it appears
that our gate could be twice as fast as the singlet qubit SWAP gate
(see section \ref{sub:Single-Qubit-Gates}), $[\sum J_{ij}]_{\textrm{avg}}$
is likely to be at least halved by the adiabatic pulses that are required
for high fidelity operation (discussed in section \ref{sec:Sources-of-Error}).
For a more experimentally achievable linear arrangement ($J_{23}=J$,
$J_{14}=0$) using an adiabatic pulse, our gate would have operation
times of roughly twice that of a single qubit SWAP gate with comparable
fidelity.

We can formalise this argument by appealing to perturbation theory
to further motivate the $\sigma_{z}\sigma_{z}$ coupling between the
qubits. Since $J_{14}+J_{23}\ll B_{0}$, we can consider the exchange
coupling terms ($V$) to be a perturbation to the Zeeman splitting
terms $(H_{0}$) of the Hamiltonian in equation \ref{eq:simple_hamiltonian}.
The first order perturbed Hamiltonian will then be $H^{1}=H_{0}+PVP$,
where $P$ is a projector onto the spin basis; and hence only the
diagonal components of $V$ can affect the eigenstates of the Hamiltonian
at first order. The exchange coupling operators are of the form $\bm{\sigma}\cdot\bm{\sigma}=\sigma_{x}\sigma_{x}+\sigma_{y}\sigma_{y}+\sigma_{z}\sigma_{z}$
(identity operators omitted), and so only the $\sigma_{z}\sigma_{z}$
terms will contribute; which when projected onto the logical subspace
looks like a logical $\sigma_{z}\sigma_{z}$ interaction. The $\sigma_{x}\sigma_{x}$
and $\sigma_{y}\sigma_{y}$ components give rise to corrections at
higher orders of perturbation theory; which nevertheless turn out
to be correctable using single qubit phase gates.

Due to the simplicity of the model, we can in fact solve the system
exactly for the eigenvalues and eigenstates by breaking the system
down into a series of two-level systems; for example, for $J_{14}\ne0$,
the two level system of $\ket{\uparrow\downarrow,\uparrow\downarrow}$
and $\ket{\downarrow\downarrow,\uparrow\uparrow}$. The energies for
all eigenstates are tabulated in the supplementary material. Since
we are principally interested in the dynamics of the logical subspace,
we restrict our attention to the states which adiabatically transform
to the logical basis $\{\ket{\uparrow\downarrow,\uparrow\downarrow},\ket{\uparrow\downarrow,\downarrow\uparrow},\ket{\downarrow\uparrow,\uparrow\downarrow},\ket{\downarrow\uparrow,\downarrow\uparrow}\}$
which we label $\{\ket{1}\ket{1},\ket{1}\ket{0},\ket{0}\ket{1},\ket{0}\ket{0}\}$
respectively. We can then write a Hamiltonian, termed the ``effective''
Hamiltonian, that reproduces the instantaneous energy spectrum. Written
in terms of effective Pauli operators $\tilde{\sigma}_{z}^{i}$ for
qubit $i$, e.g. $\tilde{\sigma}_{z}^{1}=(\ket{1}\bra{1}-\ket{0}\bra{0})\otimes I$,
the resulting effective Hamiltonian is:
\begin{eqnarray}
H_{\mathrm{eff}} & = & (\mu\Delta_{12}+\bar{B})\tilde{\sigma}_{z}^{1}+(\mu\Delta_{34}+\bar{B})\tilde{\sigma}_{z}^{2}\nonumber \\
 &  & +\frac{1}{4}(J_{14}+J_{23})\left(\tilde{\sigma}_{z}^{1}\tilde{\sigma}_{z}^{2}-II\right),\label{eq:logical_subspace_hamiltonian}
\end{eqnarray}
with $\Delta_{ij}=B_{i}-B_{j}$ and $\bar{B}$ an effective global
intra-qubit magnetic field that depends on $J_{ij}$ and the magnetic
field gradients between each pair of dots. For the precise form of
$\bar{B}$, refer to the supplementary material.

The effective Hamiltonian in equation (\ref{eq:logical_subspace_hamiltonian})
can be used to calculate the dynamical two-qubit phase accrued by
adiabatic evolution of our gate; in which case our gate will perform
a perfect CPHASE gate using a single exchange coupling pulse in a
time $\tau=\pi/(J_{14}+J_{23})_{\textrm{avg}}$ modulo known correctable
single qubit gates. One can either keep track of the single-qubit
errors described by this Hamiltonian and later correct them after
one or several gate operations, or correct them during the gate operation
by various pulse sequences \cite{Klinovaja2012}.

\section{Sources of Error\label{sec:Sources-of-Error}}

Our proposed two-qubit gate will suffer from two main sources of error:
leakage and environmental noise. Leakage from the logical subspace
will occur due to excitations to the the non-logical $\sum\hat{S}^{z}=0$
states during the course of our gate, which we suppress in our proposal
using a magnetic field gradient and adiabatic exchange pulses. While
the basic mechanics of our gate are agnostic about the details of
implementation, the nature of environmental noise depends very much
on these details. In order to make quantitative predictions about
the performance of our gate, we have chosen to mimic the noisy environment
of GaAs/AlGaAs semiconductor systems. In these systems, we anticipate
that charge fluctuations are likely to be the largest source of environmental
noise; as was observed in single qubit singlet-triplet exchange experiments
\cite{Dial2013}. As a result, in this work we neglect the Overhauser
field due to the bath of nuclear spins in the semiconductor lattice,
which should be less significant than charge noise over the time-scale
of a single gate, and which can in any case be suppressed, for example,
by nuclear state preparation \cite{Reilly2008,Foletti2009}. We also
neglect the influence of spin-flip processes arising from spin orbit
coupling\cite{Klinovaja2012}, which have been shown to occur on millisecond
time-scales \cite{Coish2007,Johnson2005} rather than the nanosecond
time-scales in which we are interested.

\subsection{Leakage\label{sub:Leakage}}

\begin{figure}
\subfloat{\includegraphics[width=1\columnwidth]{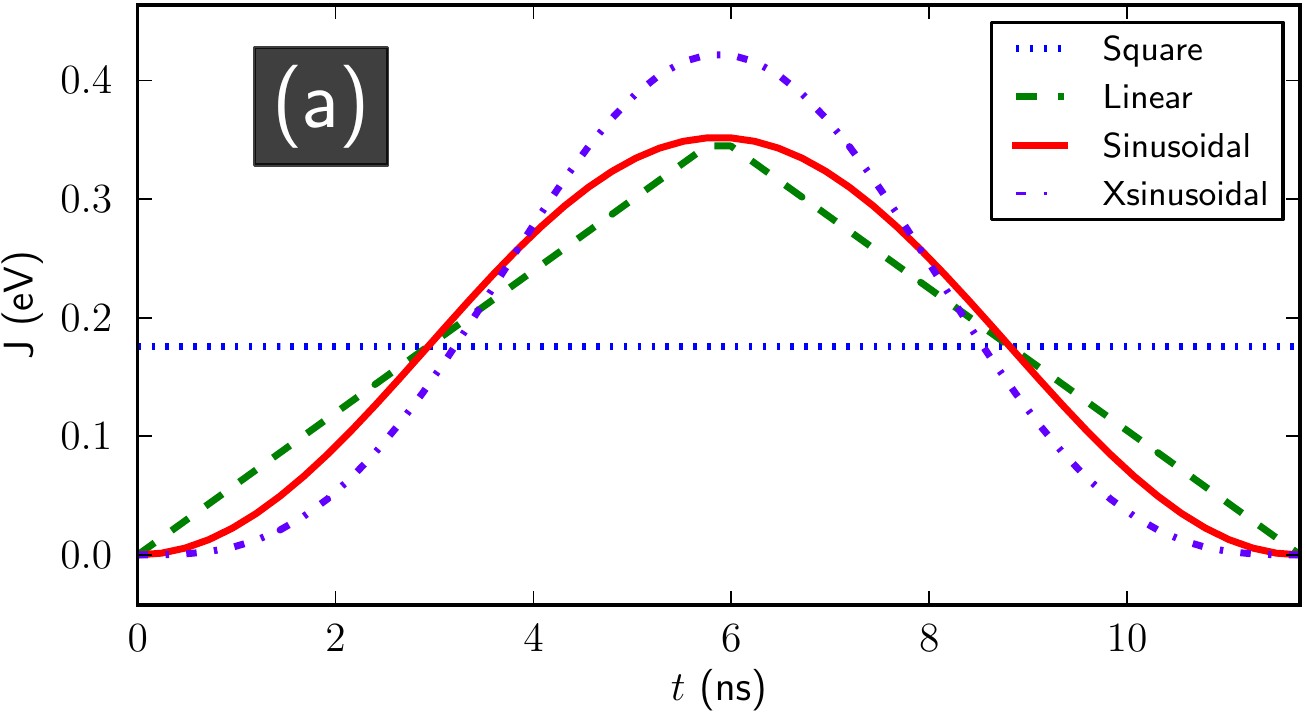}

\label{fig:pulse-profiles}}

\subfloat{\includegraphics[width=1\columnwidth]{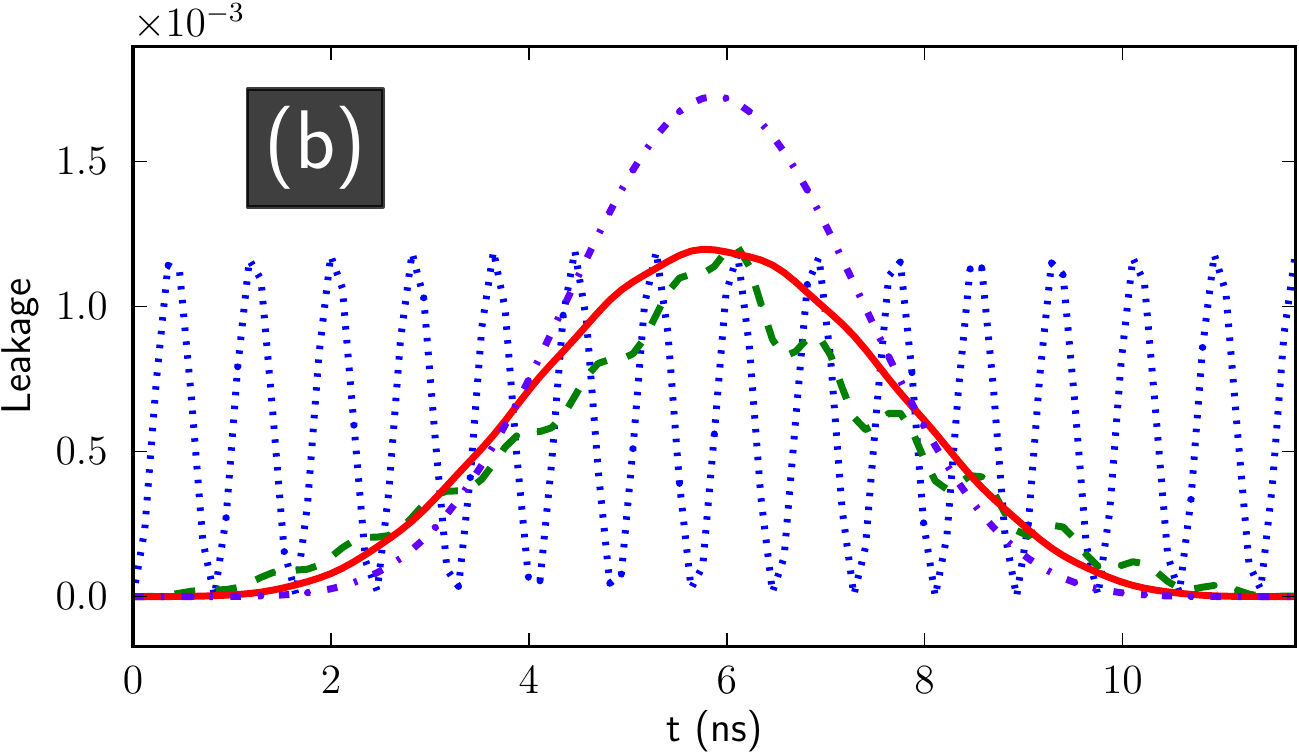}

\label{fig:profile-leakage-errors}}

\protect\caption{(colour online) a) The square, linear, sinusoidal and xsinusoidal
adiabatic pulse profiles discussed in the text. Amplitudes are chosen
to preserve gate operation time, and hence the average value of $J$.
We have here chosen $J_{\mathrm{avg}}=0.18\,\mu eV$, which corresponds
to a gate time of roughly $11.5\, ns$. b) Leakage from the logical
subspace during the operation of the gate for each of the square,
linear, sinusoidal and xsinusoidal pulses in (a). Note that the use
of adiabatic profiles can significantly reduce leakage errors at the
end of the gate operation.}
\end{figure}

We define leakage ($\mathcal{L}$) to be the probability that the
state of the system, if measured, would not be in the logical subspace:
$\mathcal{L}=1-\bra{\psi}P\ket{\psi}$, where $\psi$ is the state
of the system, and $P$ is the projector onto the logical subspace.
In the analysis of our two-qubit gate in section \ref{sub:Our-Two-Qubit},
we restricted the domain of attention to the logical subspace; explicitly
neglecting leakage. Without spin orbit coupling, leakage can only
occur to other states in the $\sum\hat{S}^{z}=0$ subspace. Although
leakage to the off-subspace states $\left\{ \ket{\uparrow\uparrow,\downarrow\downarrow},\ket{\downarrow\downarrow,\uparrow\uparrow}\right\} $
is suppressed by the energy gap $\mu\Delta B$ introduced by $\Delta B$,
if the exchange coupling terms are too quickly varied, diabatic transitions
will still occur and result in leakage probability oscillations with
frequency $\sim\mu\Delta B/h$; as seen for the square (non-adiabatic)
profile in figure \ref{fig:profile-leakage-errors}.
 Since the leakage is periodically returning to zero, it is in principle
possible minimise leakage by using precise timing of the gate in a
manner similar to Levy's original proposal\cite{Levy2002}. However,
this only works in the absence of other sources of noise, and is in
practice very difficult in any case; and we suggest that the suppression
of leakage using adiabatic pulses is substantially more robust.

An adiabatic pulse is one that turns on slowly and smoothly enough
that the system remains in an instantaneous eigenstate. The rate at
which a pulse can be turned on while remaining adiabatic depends on
energy gap between the occupied eigenstates and their neighbours.
In our case, two of the four logical states $\left\{ \ket{\uparrow\downarrow,\uparrow\downarrow},\ket{\downarrow\uparrow,\downarrow\uparrow}\right\} $
can leak to the states $\left\{ \ket{\uparrow\uparrow,\downarrow\downarrow},\ket{\downarrow\downarrow,\uparrow\uparrow}\right\} $,
which are separated in energy by approximately $\mu\Delta B$.

While a pulse can never be perfectly adiabatic, even a very simple
adiabatic pulse can greatly improve gate performance. In this paper,
we have chosen to demonstrate the behaviour of adiabatic pulses using
three representatives: linear, sinusoidal and ``xsinusoidal'', which
we compared to the non-adiabatic square pulse; as defined below:
\begin{eqnarray*}
J_{\textrm{sq}}(t) & = & J_{\textrm{avg}}=\frac{\pi}{\tau}\\
J_{\textrm{lin}}(t) & = & 2J_{\textrm{avg}}\left(1-\left|\frac{2t}{\tau}-1\right|\right)\\
J_{\textrm{sin}}(t) & = & J_{\textrm{avg}}\left(1-\cos\left(\frac{2\pi t}{\tau}\right)\right)\\
J_{\textrm{xsin}}(t) & = & J_{\textrm{avg}}\frac{6\pi^{2}}{(\pi^{2}+3)}\frac{t(\tau-t)}{\tau^{2}}\left(1-\cos\left(\frac{2\pi t}{\tau}\right)\right),
\end{eqnarray*}
each of which is depicted in figure \ref{fig:pulse-profiles}. For
ease of comparison, each profile has been normalised such that for
any given gate time $\tau$ the area is the same as a square pulse
of coupling strength $J_{\textrm{avg}}=\pi/\tau$; which in turn will
have an area $\pi$ in order to enact our gate (see section \ref{sub:Our-Two-Qubit}).
The benefits of using an adiabatic pulse are evident in figure \ref{fig:profile-leakage-errors},
in which leakage is reduced by several orders of magnitude at the
end of the gate for all adiabatic pulse profiles.

\begin{figure}
\includegraphics[width=1\columnwidth]{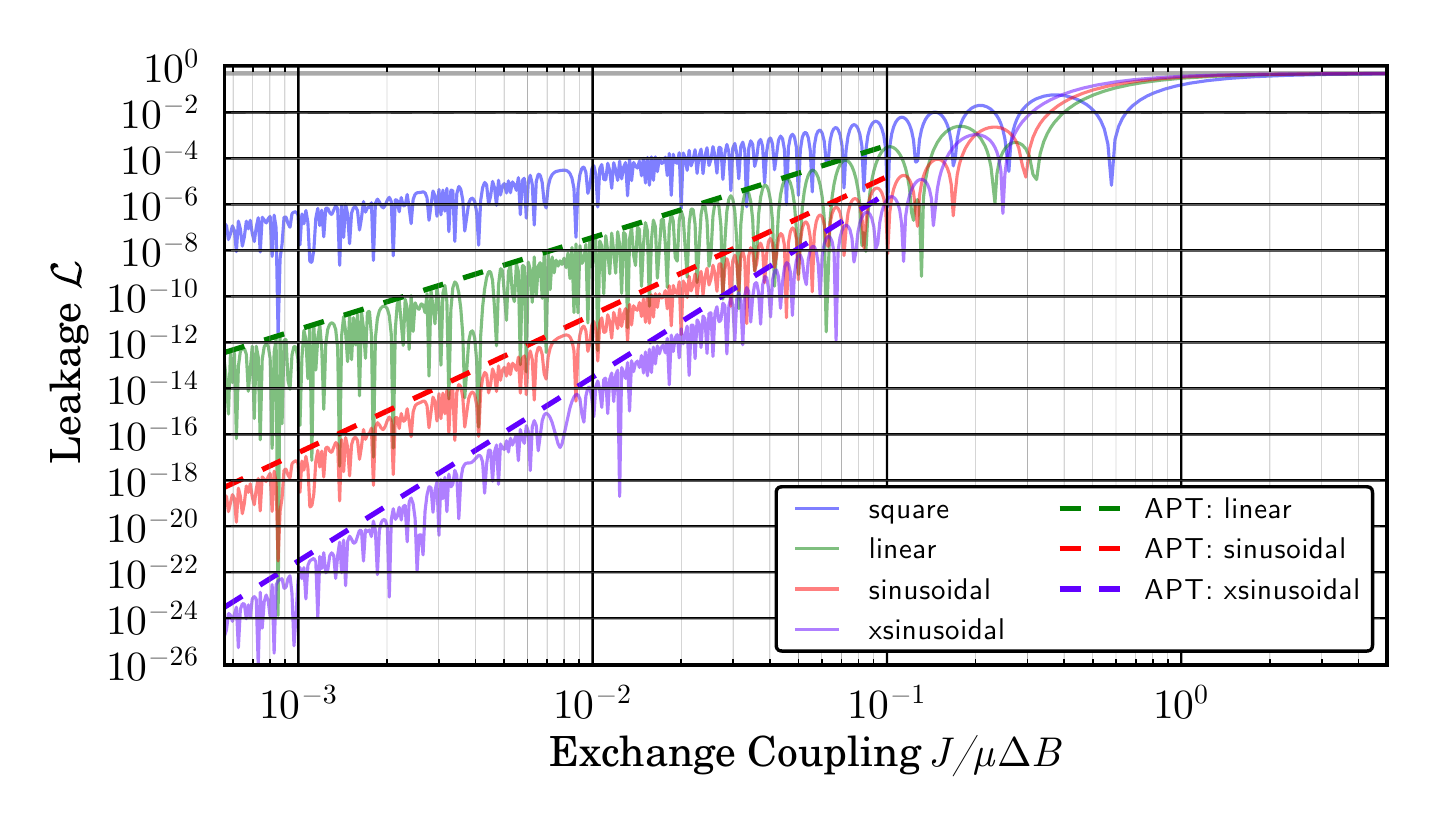}

\protect\caption{(colour online) A log-log plot of leakage immediately after a gate
operation as a function of $J/\mu\Delta B$. Upper bound predictions
from adiabatic perturbation theory (dashed lines) are compared to
data from simulations (solid traces) for the square (blue), linear
(green), sinusoidal (red) and xsinusoidal (purple) adiabatic pulses;
demonstrating reasonable agreement. \label{fig:APT behaviour}}
\end{figure}

It is possible to calculate corrections to the adiabatic approximation
that provide analytic estimates of the leakage for different pulses.
We used the adiabatic perturbation theory (APT) of de~Grandi et al.
\cite{DeGrandi2010}, which predicts that the leakage probability
scales with the lowest order derivative of the adiabatic pulse that
is discontinuous. There will always be a discontinuity at some differential
order for $t=0$ and $t=\tau$; and for the linear case, for $t=\tau/2$.
The adiabatic pulses selected for this paper were chosen such that
each profile had increasing order at which the discontinuities occurred;
and in this sense are representatives of a much larger family of adiabatic
pulses. Note too that we have avoided continuous profiles that are
not smooth, such as adiabatic ramps to a plateau; as the reductions
in adiabaticity from discontinuities would accumulate and one can
always generate a pulse with the same gate time which performs better.
For example, while we have included the linear ramp because of its
simplicity, a better choice would have been the parabola $-J_{\mathrm{avg}}\frac{6\pi}{\tau^{3}}t(t-\tau)$;
which would have avoided the larger leakage oscillations after $t=\tau/2$
visible in figure \ref{fig:profile-leakage-errors}. The corrections
arising from APT describe the amplitude and frequency of leakage oscillations,
like those seen in figure \ref{fig:profile-leakage-errors}. It is
reasonable to assume that the experimenter will not have fine-grained
temporal control due to noise and/or apparatus limitations; in which
case one wants to make the conservative assumption that the gate concludes
at a peak in these oscillations. Following de~Grandi et al. \cite{DeGrandi2010},
we calculate such a worst-case leakage probability for each of these
pulses; as shown in table \ref{tab:Maximum-leakage-error}.
\begin{table}
\begin{tabular*}{1\columnwidth}{@{\extracolsep{\fill}}ccc}
\hline
\hline
Profile & Order of Discontinuity & Maximum Leakage\tabularnewline
\hline
square & 0 & $\propto J_{avg}^{2}$ {[}for fixed $\Delta B${]}\tabularnewline
linear & 1 & $\frac{32}{\pi^{2}}\left(J_{\textrm{avg}}/\mu\Delta B\right)^{4}$\tabularnewline
sinusoidal & 2 & $16\left(J_{\textrm{avg}}/\mu\Delta B\right)^{6}$\tabularnewline
xsinusoidal & 3 & $\frac{12^{4}\pi^{2}}{4(\pi^{2}+3)^{2}}\left(J_{\textrm{avg}}/\mu\Delta B\right)^{8}$\tabularnewline
\hline
\hline
\end{tabular*}

\protect\caption{Maximum leakage error as calculated from adiabatic perturbation theory
along the lines of de~Grandi et al. \cite{DeGrandi2010} for our
selection of adiabatic pulse profiles. The order of discontinuity
refers to the lowest differential order (with respect to time) at
which the relevant pulse exhibits a discontinuity. \label{tab:Maximum-leakage-error}}
\end{table}
 These upper bounds are compared to data from our simulations in figure
\ref{fig:APT behaviour}, showing reasonable agreement. As $J$ approaches
$\mu\Delta B$ in this plot, the energy gap between the logical subspace
and the other $\sum\hat{S}^{z}=0$ states closes; causing the evolution
of the system to become strongly diabatic and resulting in a saturated
leakage of $0.5$ for all of the profiles ($0.5$ because only two
of the four logical states experience leakage from the logical subspace
under exchange). From these results, we derive a pattern whereby the
maximum leakage for a symmetric pulse with fixed area $J_{\textrm{avg}}\tau=\pi$
with first discontinuity at order $q$ will scale as $\left(J_{\textrm{avg}}/\mu\Delta B\right)^{2(q+1)}$.
The implication of this scaling is that, for any given adiabatic pulse
profile, the ratio of $J/\mu\Delta B$ gives a measure for the adiabaticity
of the adiabatic pulse.

For the rest of this paper, we assume the use of a sinusoidal adiabatic
pulse to minimise leakage. We chose the sinusoidal pulse because of
its narrow bandwidth (which may make it more straightforward to generate
in the lab) and because it boasts leakage suppression comparable with
the best in the domain likely to be of most interest ($J/\mu\Delta B\approx0.1)$.

\subsection{Charge Noise\label{sub:Charge-Noise}}

Charge noise is the result of uncontrolled electromagnetic fields
coupling into the control voltages $\varepsilon_{ij}$ of the gates
defining the quantum dots; which in turn adds noise to the exchange
couplings $J_{ij}$ of equation \ref{eq:simple_hamiltonian}. In this
section, we describe our model for charge noise and discuss its effect
on our two-qubit gate.

Since charge noise manifests itself in the control voltages $\varepsilon_{ij}$
rather than in $J_{ij}$ directly, we must find an ansatz for $J_{ij}(\varepsilon_{ij})$.
This is very difficult to do theoretically; and so we model $J$'s
dependence phenomenologically on the basis of known experimental results
on single singlet-triplet qubits (recall that singlet-triplet qubits
share the same mechanism as our two-qubit gate). In several GaAs singlet-triplet
qubit experiments \cite{Petta2005,Foletti2009,Dial2013}, an exponential
ansatz $J(\varepsilon)=J_{0}\exp(\varepsilon/\varepsilon_{D})$ with free
parameters $J_{0}$ and $\varepsilon_{D}$ has been
found to be a good phenomenological fit to experimental data over
a wide range of interesting values of $\varepsilon_{ij}$; and so
we adopt it in this work. A more complicated ansatz emerges from perturbation
theory \cite{Barthel2012}, but due to complex interations of the
electrons with the lattice in GaAs experiments, it is not clear whether
this ansatz is actually a better description.

While the noise spectrum of $\varepsilon_{ij}$ is difficult to predict
theoretically, experimental results\cite{Dial2013} suggest that the
spectrum is reasonably well approximated by a combination of low-frequency
pseudo-static components $\bar{\varepsilon}_{ij}$ that do not vary
significantly during the gate, and high frequency white-noise components
$\tilde{\varepsilon}_{ij}$ that do. We introduce a notation $\hat{\varepsilon}_{ij}$
to refer to the experimentally achieved control voltage, which includes
noise atop the theoretically desired value $\varepsilon_{ij}$; i.e.
$\hat{\varepsilon}_{ij}=\varepsilon_{ij}+\bar{\varepsilon}_{ij}+\tilde{\varepsilon}_{ij}$.
(Dial and collaborators\cite{Dial2013} also considered a power-law
charge noise spectrum which had some advantages over this model, but
we do not expect this distinction to qualitatively affect our analysis
of the gate and our simpler two-component noise spectrum allows for
more straightforward analytical analyses of gate fidelities.)

We model the pseudo-static noise component $\bar{\varepsilon}_{ij}$
to be a random variable normally distributed about zero. These low
frequency components give rise to a Gaussian decay in coherence, with
a relaxation time of $T_{2}^{*}=\sqrt{2}\left(\sigma_{\bar{\varepsilon}}\left|\frac{dJ}{d\varepsilon}\right|\right)^{-1}$,
that is reversible using spin echo pulses similar to those used in
NMR. The standard deviation of $\bar{\varepsilon}$ ($\sigma_{\bar{\varepsilon}}$)
can be determined by fitting $T_{2}^{*}$ times from free induction
decay simulations of singlet-triplet qubits to experimentally measured
values. The effect of pseudo-static noise on our gate is to shift
the average exchange coupling $J_{\textrm{avg}}$ for the gate, causing
the two-qubit phase $\int_{0}^{\tau}\left[J_{14}(\hat{\varepsilon}_{14}(t))+J_{23}(\hat{\varepsilon}_{23}(t))\right]dt$
to deviate from its ideal value of $\pi$. With our choice of ansatz
for $J$, we find that for any sampled value of $\bar{\varepsilon}$
the two-qubit phase scales like $\pi\exp(\bar{\varepsilon}/\varepsilon_{D})$.
This provides the intuition that pseudo-static noise causes an under-
or over- accrual of two-qubit phase that is approximately independent
of gate time.

The high frequency noise component $\tilde{\varepsilon}_{ij}$ is
modelled as Gaussian white noise with mean zero, which means that
it is delta correlated $\expect{\tilde{\varepsilon}_{ij}(t_{1})\tilde{\varepsilon}_{ij}(t_{2})}=D_{ij}\delta(t_{1}-t_{2})$;
where $D_{ij}$ is the spectral density of charge fluctuations. The
first order correction to the exchange terms in equation \ref{eq:simple_hamiltonian}
due to high frequency noise is $(dJ_{ij}/d\varepsilon_{ij})|_{\varepsilon_{ij}}\tilde{\varepsilon}_{ij}(\bm{\sigma}_{i}\cdot\bm{\sigma}_{j}-II)$.
Standard methods can be used to describe the average evolution of
the system in this kind of white noise in terms of a master equation
\cite{WisemanMilburn2010}. The resulting Lindblad master equation
for the system state $\rho$ is found to be:
\begin{equation}
\dot{\rho}=-i\left[H,\rho\right]+\sum_{\left<i,j\right>}D_{ij}\left|dJ_{ij}/d\varepsilon_{ij}\right|^{2}\mathcal{D}\left[\bm{\sigma}_{i}\cdot\bm{\sigma}_{j}\right],\label{eq:lindblad_master_equation}
\end{equation}
where $\mathcal{D}\left[O\right]=O^{\dagger}\rho O-\frac{1}{2}\left(\rho O^{\dagger}O+O^{\dagger}O\rho\right)$
is the Lindblad superoperator for some operator $O$. This model describes
an irreversible exponential decay in coherence, with a relaxation
time of $T_{2}=\left(8D\left|\frac{dJ}{d\varepsilon}\right|^{2}\right)^{-1}$.
The spectral density of charge fluctuations $D_{ij}$ can be determined
by fitting the results of simulations to experimental $T_{2}$ relaxation
times; for example, in Hahn echo experiments.

While this model of charge noise is very simple; we believe it captures
the essential details of the noise to which our gate is likely to
be most subjected in GaAs semiconductor systems. The model has the
nice feature of being completely specified by experimental measurements
of $T_{2}$ and $T_{2}^{*}$. An alternative would be to perform simulations
with various specified power-law noise spectra; for example, $1/f$
noise\cite{Paladino2013}; however, as noted above, we would not expect
qualitatively different results from this approach, at least for the
performance of a single gate operation.

\section{Quantifying Gate Performance\label{sec:Quantifying-Gate-Performance}}

We would like to be able to say something about the performance of
our two-qubit gate and its resilience against the sources of error
introduced in the previous section; which requires us to have a measure
of the gate's performance. While we have already used leakage as a
performance indicator in section \ref{sub:Leakage}, it neither characterises
the behaviour of the gate on the logical subspace nor includes the
effects of charge noise; and so low leakage does not imply that the
desired logical operations have actually occurred. We therefore choose
to quantify the performance of our gate by comparing the state of
the two-qubit system to some computed ideal state using the entanglement
fidelity\cite{NielsenChuang2004}. The entanglement fidelity is designed
to determine whether a gate is accurate for all possible inputs and
whether it preserves any initial entanglement with the rest of an
imagined quantum computer. It is defined in terms of a thought experiment
in which one wants to enact an ideal gate $\bar{U}$ on one half of
a maximally entangled state $\ket{\Psi_{0}}=\frac{1}{\sqrt{d}}\sum_{i}\ket{\psi_{i}}\otimes\ket{\psi_{i}}$
(where $d$ is the number of logical states, and $d=4$ for a two-qubit
gate) to yield $\ket{\bar{\Psi}}=(\bar{U}\otimes I)\ket{\Psi_{0}}$.
If instead we succeed only in performing $U$, then we yield $\ket{\Psi}=(U\otimes I)\ket{\Psi_{0}}$.
The entanglement fidelity of $U$ is then just the fidelity between
these two states:
\[
\mathcal{F}=\left|\inner{\bar{\Psi}}{\Psi}\right|^{2}.
\]
The entanglement fidelity is one if and only if the gate is perfectly
implemented, and is less than one for all other operations; with lower
values reflecting less accurate implementations of the gate. Among
the attractive features of the entanglement fidelity is its simplicity
and its close relationship with other measures of gate performance,
such as average fidelity \cite{Horodecki1999,Nielsen2002}. Notice
that our measure of fidelity assumes that the initial state of the
computation is in the logical subspace (and hence $d=4$, rather than
$d=6$). Leakage from the logical subspace will result in $\ket{\Psi}$
living in a larger space, and hence a reduction of entanglement fidelity.
The entanglement fidelity can be trivially extended to describe the
fidelity of non-unitary (noisy) implementations of the gate.

We are interested only in the performance of the two qubit component
of the implemented unitary. Since universal single qubit operations
are already possible (see section \ref{sub:Single-Qubit-Gates}) and
well characterised \cite{Petta2005}, we need only a non-trivial two
qubit gate to generate a universal set of gates for quantum computation
\cite{Lloyd1995}. Thus, while in practice our protocol produces a
CPHASE gate along with some known single qubit rotations (as described
in section \ref{sub:Our-Two-Qubit}), it would not be necessary in
some experimental implementations to correct them immediately after
each gate application; and if it were, these gates can be echoed away
by a protocol such as the one in Klinovaja et al.\cite{Klinovaja2012}.
Consequently, we compute the entanglement fidelity of our gate assuming
that the optimal single qubit corrections have been perfectly applied.
In practice this means that we compare our gate to a constructed ideal
unitary $\bar{U}$ that maintains the ideal two-qubit phase while
also including whatever single qubit z-rotations we find in the simulation
of our gate. This corresponds to an ansatz $\bar{U}=e^{i\phi_{II}}e^{i\phi_{ZI}\tilde{\sigma}_{z}^{1}}e^{i\phi_{IZ}\tilde{\sigma}_{z}^{2}}e^{i\phi_{ZZ}\tilde{\sigma}_{z}^{1}\tilde{\sigma}_{z}^{2}}$
with global and single-qubit phases $\phi_{II}$, $\phi_{IZ}$ and $\phi_{ZI}$ extracted from simulations,
and the two-qubit phase $\phi_{ZZ}$ set to $\pi$. There are some subtleties to this
process which we discuss in the supplementary material.

The simplicity of our two singlet-triplet qubit model allows us to gain some
intuition about the entanglement fidelity by considering the analytic solution
for fidelity in terms of the leakage at the end of the gate $\mathcal{L}_{0}$
and the parameter $\Delta=\phi_{ZZ}-\bar{\phi}_{ZZ}$, where $\phi_{ZZ}$ and
$\bar{\phi}_{ZZ}=\pi$ are the extracted and ideal two qubit phases respectively.
Note that $\Delta$ characterises any under- or over- accrual of two-qubit phase
acquired during the gate operation. We show in the supplementary material
that the entanglement fidelity $\mathcal{F}$ at the end of the gate is:
\begin{align}
\mathcal{F} & = \frac{1}{2}\left(1+\sqrt{1-2\mathcal{L}_{0}}\cos\left(2\Delta\right)-\mathcal{L}_{0}\right)\\
 & = 1-\mathcal{L}_{0}-\Delta^{2}+\mathcal{O}(\mathcal{L}_{0}^{2})+\mathcal{O}(\mathcal{L}_{0}\Delta^{2})+\mathcal{O}(\Delta^{4}).
\end{align}
Perfect gate implementations will have $\Delta=0$, whereas in the
presence of charge noise it will assume non-zero values since $\phi_{ZZ}\simeq\int_{0}^{\tau}\left[J_{14}(t)+J_{23}(t)\right]dt$.

Our charge noise model of section \ref{sub:Charge-Noise} allows us
to make some more quantitative statements about $\Delta$. Consider
a square pulse for simplicity (adiabatic pulses are not expected to
lead to qualitatively different results) with first order noise perturbations,
$J(t)=J_{\textrm{avg}}+(\bar{\varepsilon}+\tilde{\varepsilon})(dJ/d\varepsilon)|_{J(\varepsilon)=J_{\mathrm{avg}}}$
where $\bar{\varepsilon}$ and $\tilde{\varepsilon}$ are the pseudo-static
and high frequency potential fluctuations caused by charge noise (see
section \ref{sub:Charge-Noise}). The phase error $\Delta$ is then:
\[
\Delta=\frac{\pi}{J_{\mathrm{avg}}}\left.\frac{dJ}{d\varepsilon}\right|_{J=J_{\mathrm{avg}}}(\bar{\varepsilon}+\tilde{\varepsilon})_{\mathrm{avg}}
\]
where $(\bar{\varepsilon}+\tilde{\varepsilon})_{\mathrm{avg}}=\int_{0}^{\tau}\left[(\bar{\varepsilon}+\tilde{\varepsilon})/\tau\right]dt$
is the average charge fluctuation during the operation of the gate.
Our noise model posits that $\bar{\varepsilon}$ and $\tilde{\varepsilon}$
are both independent and Gaussian distributed; which allows us to
easily calculate the statistical properties of $(\bar{\varepsilon}+\tilde{\varepsilon})_{\mathrm{avg}}$:
$\expect{(\bar{\varepsilon}+\tilde{\varepsilon})_{\mathrm{avg}}}=0$
and $\expect{(\bar{\varepsilon}+\tilde{\varepsilon})_{\mathrm{avg}}^{2}}=\sigma_{\bar{\varepsilon}}^{2}+D/\tau$.
While different charge spectra beyond our model would lead to different
time dependences for $\expect{(\bar{\varepsilon}+\tilde{\varepsilon})_{\mathrm{avg}}^{2}}$,
it would always be qualitatively true that non-pseudo-static noise
averages out over long enough time scales. This implies the statistical
properties of $\Delta$:
\begin{eqnarray*}
\expect{\Delta} & = & 0\\
\expect{\Delta^{2}} & = & \frac{\pi^{2}}{J_{\mathrm{avg}}^{2}}\left.\frac{dJ}{d\varepsilon}\right|_{J=J_{\mathrm{avg}}}^{2}\left(\sigma_{\bar{\varepsilon}}^{2}+D/\tau\right).
\end{eqnarray*}

If we further assume the exponential ansatz $J\simeq J_{0}\exp(\varepsilon/\varepsilon_{D})$
we obtain a particularly simple estimate of the expected entanglement
fidelity $\expect{\mathcal{F}}$:
\begin{equation}
\left\langle \mathcal{F}\right\rangle \simeq1-\mathcal{L}_{0}-\pi^{2}\frac{\sigma_{\bar{\varepsilon}}^{2}+D/\tau}{\varepsilon_{D}^{2}}.\label{eq:fidelity-relation}
\end{equation}

This formula affords us the important intuition that one is in principle
able to maximise the fidelity of our gate by choosing sufficiently
long gate times; at which point fidelity will be limited by a pseudo-static
noise floor that also determines the fidelity of single-qubit operations.
This is evident because both leakage and the effect of high frequency
noise are monotonically decreasing functions of the choice of gate time;
and pseudo-static noise contributions are constant. The leakage $\mathcal{L}_{0}$
due to the pulse scales as $(\tau\Delta B)^{-c}$ for some $c$ that
depends on the adiabaticity of the pulse (as shown in section \ref{sub:Leakage}).
Additional leakage contributions arise from high frequency noise,
at a rate $4D|dJ/d\varepsilon|^{2}$ predicted by the master equation
(\ref{eq:lindblad_master_equation}), and hence a contribution to
$\mathcal{L}_{0}$ proportional to $D/(\tau\varepsilon_{D}^{2})$.
Since all sources of fidelity dimunition apart from pseudo-static
noise decrease as gate time increases, pseudo-static noise will dominate
at sufficiently long gate times; after which the fidelity becomes
roughly independent of gate time. In the limit that leakage is no
longer the dominant noise contribution, equation \ref{eq:fidelity-relation}
reduces to the fidelity relation for a single qubit gate; implying
that our gate would operate with essentially identical fidelity as
single qubit operations.

Although we have largely neglected the effects of Overhauser field
fluctuations on our simulations, since they are expected to be less
significant than charge noise in the usual regime of operation, it
would be straightforward to include them in this kind of analysis.
The main effect of these fluctuations is to implement random single
qubit unitaries during the gate; which would result in a reduction
of gate fidelity proportional to the variance of the field fluctuation.

\section{Simulations\label{sec:Simulations}}

\begin{figure*}
\subfloat{\label{fig:sinusoidal-leakage-noiseless}}\subfloat{\label{fig:sinusoidal-fidelity-noiseless}}\subfloat{\label{fig:sinusoidal-leakage-noisy}}\subfloat{\label{fig:sinusoidal-fidelity-noisy}}
\includegraphics[width=1\textwidth]{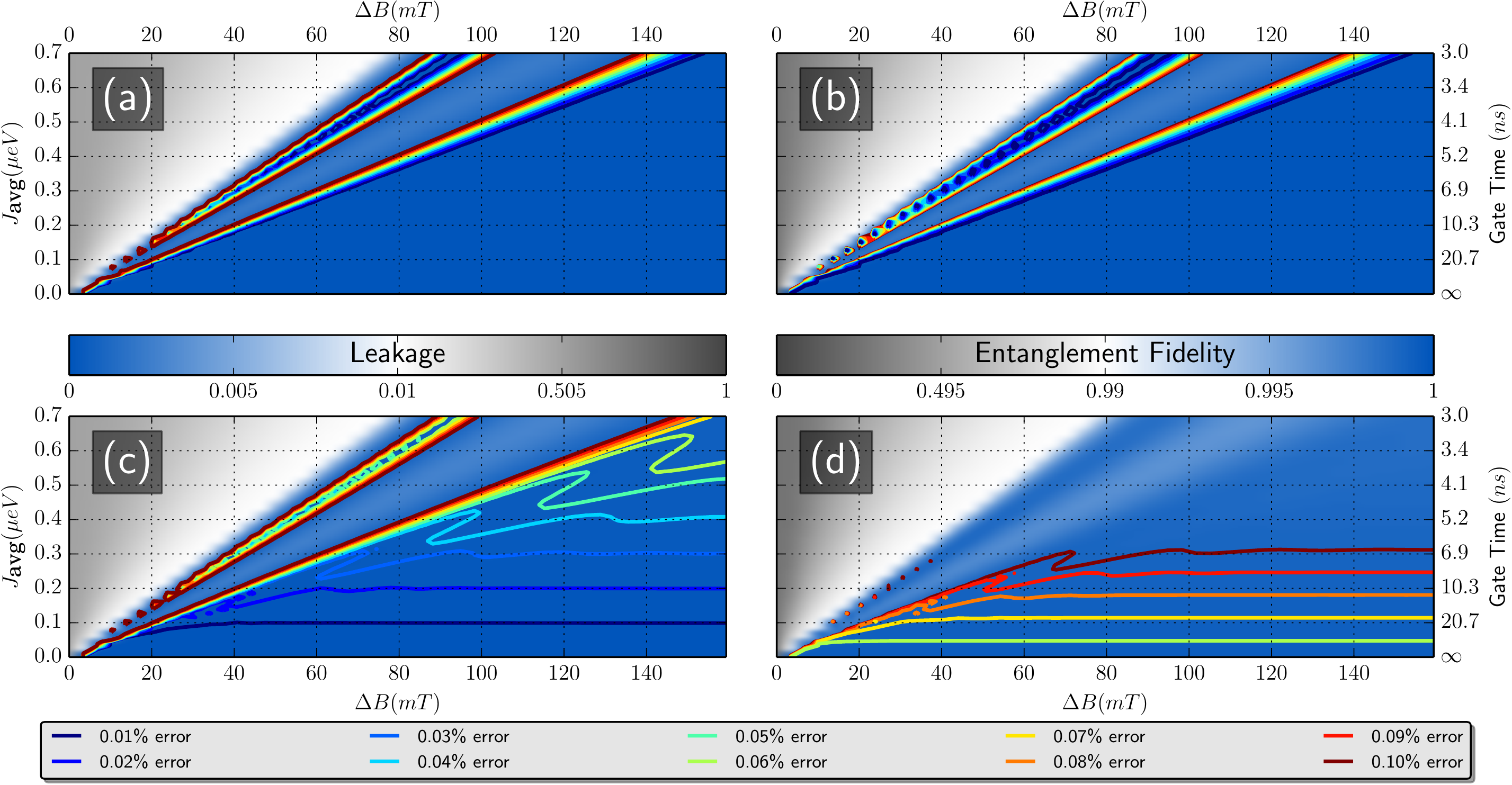}
\protect\caption{(colour online) Simulated leakage and entanglement fidelity immediately
after a single two-qubit gate operation as a function of $J_{\mathrm{avg}}$
and $\Delta B$; with gate time corresponding to each labelled $J_{\mathrm{avg}}$
shown on the far right axes. Subplots (a) and (c) show the leakage
results with and without charge noise respectively; and subplots (b)
and (d) show the entanglement fidelity results with and without charge
noise respectively. The colour maps for leakage and entanglement fidelity
are shown in the corresponding column, and are chosen such that comparable
colours indicate comparable errors. The colour maps are white at $1\%$
error; with larger error shown in shades of grey; and lesser error
in shades of blue. Contours corresponding to 0.01\%-0.10\% error are
drawn at 0.01\% intervals. \label{fig:JvsdB plots}}
\end{figure*}

We are left now only to demonstrate the performance of our two singlet-triplet
gate in simulations. For the purposes of this section, we integrate
the time dependent Heisenberg Hamiltonian (with and without the Lindblad
terms) as described in equations \ref{eq:simple_hamiltonian} and
\ref{eq:lindblad_master_equation}. One might worry that the weak
tunnelling approximation described in section \ref{sec:Model} might
lead to appreciable errors; but simulations of a full Hubbard model
generated indistinguishable results in all of our tests. We have chosen
to consider a case where $J_{14}=0$, and henceforth $J=J_{23}$,
because we expect a linear arrangement of quantum dots to be more
accessible to experimental implementation. Simulating a square configuration
is a trivial extension that does not alter the physics; indeed it
improves the gate speed for any given leakage error, and reduces the
complexity of the unwanted single qubit gates (refer to the supplementary
material for more information). For reasons mentioned in section \ref{sub:Leakage},
we have chosen to use the sinusoidal adiabatic pulse in these simulations.

All of the parameters used in these simulations have been chosen with
current experiments in mind. The average background magnetic field
$B_{0}$ is maintained as the dominant energy scale, and set to be
$200\, mT$. As described in section \ref{sub:Charge-Noise}, simulations
of our gate require us to choose an ansatz for $J(\varepsilon$);
the details of which are not important provided that the dependence
matches closely phenomenological results. In these simulations we
use $J(\varepsilon)=J_{0}\exp(\varepsilon/\varepsilon_{D})$, with
the free parameters $J_{0}$ and $\varepsilon_{D}$ chosen to
roughly match the experimental results of Dial and collaborators \cite{Dial2013}.
In simulations we add a very small negative constant offset to allow
for $J(\varepsilon)$ to be exactly zero, which does not significantly
affect the results. Also following the prescription of section \ref{sub:Charge-Noise},
we calibrated the noise parameters $\sigma_{\bar{\varepsilon}}$ and
$D$ such that the $T_{2}^{*}$ and $T_{2}$ times in simulations
of single-qubit gates were somewhat typical\cite{Dial2013}; specifically,
we calibrated $\sigma_{\bar{\varepsilon}}$ and $D$ such that $T_{2}^{*}\approx100\, ns$
and $T_{2}\approx1\,\mu s$. We summarise our choice of parameters
in table \ref{tab:Parameters}.
\begin{table}

\begin{tabular*}{1\columnwidth}{@{\extracolsep{\fill}}ccc}
\hline
\hline
Parameter & Value \tabularnewline
\hline

\multicolumn{1}{l}{Magnetic Field:} \tabularnewline
$B_{0}$ & $200\, mT$ \tabularnewline
\hline
\multicolumn{1}{l}{Noise:} \tabularnewline
$\sigma_{\bar{\varepsilon}}$ & $10.3\, \mu V$\tabularnewline
$D$ & $100\,\mu V^{2}\, ns$ \tabularnewline
\hline
\multicolumn{1}{l}{Exponential Ansatz:}\tabularnewline
$J_{0}$ & $-82.7\,\mu eV\,(\approx20GHz$)\tabularnewline
$\varepsilon_{D}$ & $0.35\, mV$\tabularnewline
\hline
\hline
\end{tabular*}

\protect\caption{Model parameters kept constant while exploring $J$ and $\Delta B$.
The global magnetic field is nominal. The noise parameters $\sigma_{\bar{\varepsilon}}$
and $D$ were calibrated by respectively matching somewhat typical
values of $T_{2}^{*}$ and $T_{2}$ from experiment\cite{Dial2013}.
The parameters for the exponential ansatz $J(\varepsilon)=J_{0}\exp(\varepsilon/\varepsilon_{D})$ were chosen to
roughly match the experimental results of Dial and collaborators \cite{Dial2013}.
\label{tab:Parameters}}
\end{table}

In figure \ref{fig:JvsdB plots} we plot leakage with and without
charge noise (\ref{fig:sinusoidal-leakage-noiseless} and \ref{fig:sinusoidal-leakage-noisy}
respectively), and entanglement fidelity with and without noise (\ref{fig:sinusoidal-fidelity-noiseless}
and \ref{fig:sinusoidal-fidelity-noisy} respectively), at the end
of a single two-qubit gate operation as a function of the average
exchange coupling $J_{\mathrm{avg}}$ and inter-qubit magnetic field
gradient $\Delta B$. The gate time corresponding to each value of
$J_{\mathrm{avg}}$ is labelled on the rightmost axes. We use a segmented
colour map which goes through white at 1\% error; with blue toward
0\% and dark grey toward 100\%. Contours corresponding to 0.01\%-0.10\%
error are drawn at 0.01\% intervals; corresponding to the regime in
which fault-tolerant computing starts to become feasible (typically
$10^{-2}-10^{-4}$ )\cite{Gottesman2007}. The positions of the contours
on both the leakage and fidelity plots roughly correspond, because
$\mathcal{F}\approx1-\mathcal{L}_{0}$ (see section \ref{sec:Quantifying-Gate-Performance}).

The simulations confirm several important qualities of our gate. The
radial nature of the contours from the origin indicates that our
gate's leakage is largely predicted by the ratio $J_{\mathrm{avg}}/\mu\Delta B$,
as described in section \ref{sub:Leakage}. We also observe the qualitative
features of anticipated in section \ref{sec:Quantifying-Gate-Performance}:
that charge noise causes the entanglement fidelity to decrease even
more quickly than leakage increases due to its being sensitive to
phases on the logical subspace (seen in \ref{eq:fidelity-relation});
that leakage ceases to be the dominant source of error (for any given
fidelity) at long enough gate times; and that fidelity increases approximately
hyperbolically with gate time (or equivalently, decreases linearly with
$J_{\mathrm{avg}}$).

By way of ballpark numbers, we find that our two-qubit gate has fidelities
in excess of $\sim99.9\%$ for gate times longer than around $7\, ns$
and magnetic field gradients of around $80\, mT$. In this regime,
leakage is no longer the dominant source of noise; and gate fidelities
are affected by noise in essentially the same way as single qubit
gates, as described in section \ref{sec:Quantifying-Gate-Performance}.
On this basis, we feel that our gate may be well suited for quantum
computation using singlet-triplet qubits.

\section{Discussion\label{sec:Discussion}}

In this paper we have analysed the performance of an exchange-based
two-qubit gate for singlet-triplet qubits. Our approach uses a magnetic
field gradient to suppress spin-flip transitions that would otherwise
lead to leakage errors. We have shown that adiabatic pulses can reduce
the leakage probability by several orders of magnitude. We have also
investigated the effect of charge noise on the performance of our
gate; showing that, by running the gate sufficiently slowly, it is
possible to achieve entanglement fidelities that are comparable to
those of single qubit operations. In this limit, we showed that
the performance is limited by low frequency charge noise. Two-qubit
gate simulations demonstrated that this regime can be reached using
realistic exchange couplings and magnetic field gradients.

The two-qubit gate we have described works whenever there is an effective
exchange coupling between the qubits. This could be a direct exchange
coupling (as we have envisaged here), or an indirect coupling through
an intermediate dot which has recently been shown
to generate an effective exchange interaction \cite{Mehl2014}.

The approach of energetically suppressing spin-flip transitions in
order to implement two-qubit gates using exchange coupling has application
in other qubit architectures; for example, we used a similar approach
in our proposal for two-qubit gates for the resonant-exchange qubit
\cite{Doherty2013}. We anticipate that the use of adiabatic pulses
will greatly reduce leakage in that scheme also.
\begin{acknowledgments}
We thank S. Bartlett, B. Halperin, D. Loss, C. Marcus, I. Neder, D.
Reilly, and M. Rudner for discussions. Research was supported by the
Office of the Director of National Intelligence, Intelligence Advanced
Research Projects Activity (IARPA), through the Army Research Office
grant W911NF-12-1-0354. We acknowledge support from the ARC via the
Centre of Excellence in Engineered Quantum Systems (EQuS), project
number CE110001013.
\end{acknowledgments}

\bibliographystyle{apsrev}

\onecolumngrid
\newpage
\appendix
\gdef\appendixname{Supplementary}

\section{Analytically solving for the eigen-energies of a two singlet-triplet
qubit system\label{sec:Analytically-solving-for}}

In the paper we claimed that it was possible to derive exact solutions
the eigen- energies and states of the Hamiltonian:

\begin{equation}
H(t)=\mu\sum_{n=1}^{N}B_{n}(t)\sigma_{z}^{n}+\frac{1}{4}\sum_{\left<i,j\right>}J_{ij}(t)(\bm{\sigma}^{i}\cdot\bm{\sigma}^{j}-I),\label{eq:simple_hamiltonian}
\end{equation}
where $J_{14}$ and $J_{23}$ are the only non-zero exchange couplings.
We do so in the following for the case were $J_{14}=0$ (the ``linear''
configuration), and where both $J_{14}$ and $J_{23}$ are non-zero;
which has a special case $J_{14}=J_{23}$ (the ``square'' configuration).

\subsection{Simple two-level system}

Consider first a two-level system with Hamiltonian:
\[
H=\frac{\Delta B}{2}\sigma_{z}+\frac{J}{2}\sigma_{x},
\]
where we consider the spin states to be the eigenstates of $\sigma_{z}$,
and where in principle each parameter can be time dependent. This
Hamiltonian has eigen-energies:

\begin{eqnarray*}
E_{\pm} & = & \pm\frac{\Delta B}{2}\sqrt{1+\xi^{2}},
\end{eqnarray*}
with $\xi=\frac{J}{\Delta B}$. The corresponding eigenstates are
given by:

\begin{eqnarray*}
\ket{\psi}_{+} & = & \up+(\sqrt{1+\xi^{2}}-1)/\xi\down\\
\ket{\psi}_{-} & = & \up-(\sqrt{1+\xi^{2}}+1)/\xi\down.
\end{eqnarray*}

In the following sections we will break up more complicated Hamiltonians
into two-level systems, and solve them by comparing them to these
results.

\subsection{Singlet-triplet qubit system\label{sub:Singlet-Triplet-Qubit-System}}

We derive the solutions for a singlet-triplet qubit system; before
providing solutions for two singlet-triplet systems in subsequent
sections. The Hamiltonian for single singlet-triplet qubit system
is:

\[
H=\mu B_{1}\sigma_{z}^{1}+\mu B_{2}\sigma_{z}^{2}+\frac{J}{4}(\bm{\sigma}^{1}\cdot\bm{\sigma}^{2}-I)\,.
\]

The effect of the exchange coupling term is to lower the singlet state
compared to the triplet states. The $\sum S_{z}=1$ and $\sum S_{z}=-1$
states $\ket{\uparrow\uparrow}$ and $\ket{\downarrow\downarrow}$
are unaffected by the exchange coupling; and so their eigen-energies
are $\mu(B_{1}+B_{2})$ and $-\mu(B_{1}+B_{2})$ respectively. The
effect of the Hamiltonian on the remaining singlet and triplet states
with $\sum S_{z}=0$, which corresponds to the logical subspace as
described in the main text, can be written as:
\[
\left(\begin{array}{cc}
\mu\Delta_{12}-\frac{J}{2} & \frac{J}{2}\\
\frac{J}{2} & -\mu\Delta_{12}-\frac{J}{2}
\end{array}\right)\left(\begin{array}{c}
\ket{\uparrow\downarrow}\\
\ket{\downarrow\uparrow}
\end{array}\right),
\]
with $\Delta_{12}=B_{1}-B_{2}$. By comparison with the simple two-level
system in the previous section, we find that the eigen-energies of
the states which adiabatically conform to the logical spin basis states
$\ket{\uparrow\downarrow}$ and $\ket{\downarrow\uparrow}$ in the
limit that $J=0$ are:
\begin{eqnarray*}
E_{\uparrow\downarrow} & = & \mu\Delta_{12}\sqrt{1+\xi^{2}}-\frac{J}{2}\\
E_{\downarrow\uparrow} & = & -\mu\Delta_{12}\sqrt{1+\xi^{2}}-\frac{J}{2}
\end{eqnarray*}
with $\xi=J/\mu\Delta_{12}$. The eigenstates will have the same form
as those of the previous section; where $\up\mapsto\ket{\uparrow\downarrow}$
and $\down\mapsto\ket{\downarrow\uparrow}$.

\subsection{Linear configuration for two singlet-triplet qubits}

We now consider the energy spectrum of a two singlet-triplet qubit
system in the linear configuration; that is, with exchange coupling
only between the second and third quantum dots. The Hamiltonian for
this system is given by:
\[
H=\mu\sum_{n=1}^{4}B_{n}\sigma_{z}^{n}+\frac{1}{4}J_{23}(\bm{\sigma}^{2}\cdot\bm{\sigma}^{3}-I).
\]

As in the paper, we imagine a large global magnetic field; making
the spin basis a natural one. Moreover, since we are not going to
be considering the possibility of spin-flip, only system states with
like total $\sum S_{Z}$ can communicate. Grouping the states by like
$\sum S_{z}$, we find 5 different groups with $\sum S_{z}=0,\pm1,\pm2$.
By comparision to the previous section, it is clear that only eigenstates
that have different spin states for quantum dots 2 and 3 will be coupled
by the interaction. This forms a series of two level coherences; i.e.
$\ket{\uparrow\uparrow,\downarrow\uparrow}\rightleftarrows\ket{\uparrow\downarrow,\uparrow\uparrow};\,\ket{\uparrow\uparrow,\downarrow\downarrow}\rightleftarrows\ket{\uparrow\downarrow,\uparrow\downarrow}$;
etc. We use the same magnetic field basis used in the paper: the global
background magnetic field $B_{0}=\frac{1}{4}\sum_{n}B_{n}$, the inter-qubit
gradient $\Delta B=\frac{1}{2}\left(B_{1}+B_{2}-B_{3}-B_{4}\right)$,
and the intra-qubit gradients $\Delta_{12}$ and $\Delta_{34}$ with
$\Delta_{ij}=B_{i}-B_{j}$.

By applying a similar analysis as that in the previous section, we
tabulate the energies for the states which adiabatically conform to
the spin states below.

\[
\begin{array}{ccc}
\mathbf{m_{z}} & \textbf{Eigenstate} & \textbf{Energy}\\
2 & \ket{\uparrow\uparrow\uparrow\uparrow} & 4\mu B_{0}\\
1 & \ket{\uparrow\uparrow\uparrow\downarrow} & 2\mu B_{0}+\mu\Delta B+\mu\Delta_{34}\\
1 & \ket{\uparrow\uparrow\downarrow\uparrow} & 2\mu B_{0}+\frac{1}{2}\mu(\Delta_{12}-\Delta_{34})-\frac{1}{2}J_{23}+\sqrt{\left(\mu\Delta B-\frac{1}{2}\mu(\Delta_{12}+\Delta_{34})\right)^{2}+\frac{1}{4}J_{23}^{2}}\\
1 & \ket{\uparrow\downarrow\uparrow\uparrow} & 2\mu B_{0}+\frac{1}{2}\mu(\Delta_{12}-\Delta_{34})-\frac{1}{2}J_{23}-\sqrt{\left(\mu\Delta B-\frac{1}{2}\mu(\Delta_{12}+\Delta_{34})\right)^{2}+\frac{1}{4}J_{23}^{2}}\\
1 & \ket{\downarrow\uparrow\uparrow\uparrow} & 2\mu B_{0}-\mu\Delta B-\mu\Delta_{12}\\
0 & \ket{\uparrow\uparrow\downarrow\downarrow} & \mu\Delta B+\frac{1}{2}\mu(\Delta_{12}+\Delta_{34})-\frac{1}{2}J_{23}+\sqrt{\left(\mu\Delta B-\frac{1}{2}\mu(\Delta_{12}+\Delta_{34})\right)^{2}+\frac{1}{4}J_{23}^{2}}\\
0 & \ket{\uparrow\downarrow\uparrow\downarrow} & \mu\Delta B+\frac{1}{2}\mu(\Delta_{12}+\Delta_{34})-\frac{1}{2}J_{23}-\sqrt{\left(\mu\Delta B-\frac{1}{2}\mu(\Delta_{12}+\Delta_{34})\right)^{2}+\frac{1}{4}J_{23}^{2}}\\
0 & \ket{\uparrow\downarrow\downarrow\uparrow} & \mu\Delta_{12}-\mu\Delta_{34}\\
0 & \ket{\downarrow\uparrow\uparrow\downarrow} & -\mu\Delta_{12}+\mu\Delta_{34}\\
0 & \ket{\downarrow\uparrow\downarrow\uparrow} & -\mu\Delta B-\frac{1}{2}\mu(\Delta_{12}+\Delta_{34})-\frac{1}{2}\, J_{23}+\sqrt{\left(\mu\Delta B-\frac{1}{2}\mu(\Delta_{12}+\Delta_{34})\right)^{2}+\frac{1}{4}J_{23}^{2}}\\
0 & \ket{\downarrow\downarrow\uparrow\uparrow} & -\mu\Delta B-\frac{1}{2}\mu(\Delta_{12}+\Delta_{34})-\frac{1}{2}\, J_{23}-\sqrt{\left(\mu\Delta B-\frac{1}{2}\mu(\Delta_{12}+\Delta_{34})\right)^{2}+\frac{1}{4}J_{23}^{2}}\\
-1 & \ket{\uparrow\downarrow\downarrow\downarrow} & -2\mu B_{0}+\mu\Delta B+\mu\Delta_{12}\\
-1 & \ket{\downarrow\uparrow\downarrow\downarrow} & -2\mu B_{0}-\frac{1}{2}\mu(\Delta_{12}-\Delta_{34})-\frac{1}{2}\, J_{23}+\sqrt{\left(\mu\Delta B-\frac{1}{2}\mu(\Delta_{12}+\Delta_{34})\right)^{2}+\frac{1}{4}J_{23}^{2}}\\
-1 & \ket{\downarrow\downarrow\uparrow\downarrow} & -2\mu B_{0}-\frac{1}{2}\mu(\Delta_{12}-\Delta_{34})-\frac{1}{2}\, J_{23}-\sqrt{\left(\mu\Delta B-\frac{1}{2}\mu(\Delta_{12}+\Delta_{34})\right)^{2}+\frac{1}{4}J_{23}^{2}}\\
-1 & \ket{\downarrow\downarrow\downarrow\uparrow} & -2\mu B_{0}-\mu\Delta B-\mu\Delta_{34}\\
-2 & \ket{\downarrow\downarrow\downarrow\downarrow} & -4\mu B_{0}
\end{array}
\]

\subsection{Non-linear configuration for two singlet-triplet qubits}

We now consider the more general configuration in which both $J_{14}$
and $J_{23}$ are non-zero; and present an analogous table of eigenvalues
(as above). In the case where $J_{14}=J_{23}$ we form what we call
the square configuration. Since the two exchange couplings are acting
identically in a disjoint system (since $J_{12}$ and $J_{34}$ are
turned off); these eigenvalues follow immediately from those above.

\[
\begin{array}{ccc}
\mathbf{m_{z}} & \textbf{Eigenstate} & \textbf{Energy}\\
2 & \ket{\uparrow\uparrow\uparrow\uparrow} & 4\mu B_{0}\\
1 & \ket{\uparrow\uparrow\uparrow\downarrow} & 2\mu B_{0}-\frac{1}{2}\mu(\Delta_{12}-\Delta_{34})-\frac{1}{2}J_{14}+\sqrt{\left(\mu\Delta B+\frac{1}{2}\mu(\Delta_{12}+\Delta_{34})\right)^{2}+\frac{1}{4}J_{14}^{2}}\\
1 & \ket{\uparrow\uparrow\downarrow\uparrow} & 2\mu B_{0}+\frac{1}{2}\mu(\Delta_{12}-\Delta_{34})-\frac{1}{2}J_{23}+\sqrt{\left(\mu\Delta B-\frac{1}{2}\mu(\Delta_{12}+\Delta_{34})\right)^{2}+\frac{1}{4}J_{23}^{2}}\\
1 & \ket{\uparrow\downarrow\uparrow\uparrow} & 2\mu B_{0}+\frac{1}{2}\mu(\Delta_{12}-\Delta_{34})-\frac{1}{2}J_{23}-\sqrt{\left(\mu\Delta B-\frac{1}{2}\mu(\Delta_{12}+\Delta_{34})\right)^{2}+\frac{1}{4}J_{23}^{2}}\\
1 & \ket{\downarrow\uparrow\uparrow\uparrow} & 2\mu B_{0}-\frac{1}{2}\mu(\Delta_{12}-\Delta_{34})-\frac{1}{2}J_{14}-\sqrt{\left(\mu\Delta B+\frac{1}{2}\mu(\Delta_{12}+\Delta_{34})\right)^{2}+\frac{1}{4}J_{14}^{2}}\\
0 & \ket{\uparrow\uparrow\downarrow\downarrow} & -\frac{1}{2}(J_{14}+J_{23})+\sqrt{\left(\mu\Delta B+\frac{1}{2}\mu(\Delta_{12}+\Delta_{34})\right)^{2}+\frac{1}{4}J_{14}^{2}}+\sqrt{\left(\mu\Delta B-\frac{1}{2}\mu(\Delta_{12}+\Delta_{34})\right)^{2}+\frac{1}{4}J_{23}^{2}}\\
0 & \ket{\uparrow\downarrow\uparrow\downarrow} & -\frac{1}{2}(J_{14}+J_{23})+\sqrt{\left(\mu\Delta B+\frac{1}{2}\mu(\Delta_{12}+\Delta_{34})\right)^{2}+\frac{1}{4}J_{14}^{2}}-\sqrt{\left(\mu\Delta B-\frac{1}{2}\mu(\Delta_{12}+\Delta_{34})\right)^{2}+\frac{1}{4}J_{23}^{2}}\\
0 & \ket{\uparrow\downarrow\downarrow\uparrow} & \mu\Delta_{12}-\mu\Delta_{34}\\
0 & \ket{\downarrow\uparrow\uparrow\downarrow} & -\mu\Delta_{12}+\mu\Delta_{34}\\
0 & \ket{\downarrow\uparrow\downarrow\uparrow} & -\frac{1}{2}(J_{14}+J_{23})-\sqrt{\left(\mu\Delta B+\frac{1}{2}\mu(\Delta_{12}+\Delta_{34})\right)^{2}+\frac{1}{4}J_{14}^{2}}+\sqrt{\left(\mu\Delta B-\frac{1}{2}\mu(\Delta_{12}+\Delta_{34})\right)^{2}+\frac{1}{4}J_{23}^{2}}\\
0 & \ket{\downarrow\downarrow\uparrow\uparrow} & -\frac{1}{2}(J_{14}+J_{23})-\sqrt{\left(\mu\Delta B+\frac{1}{2}\mu(\Delta_{12}+\Delta_{34})\right)^{2}+\frac{1}{4}J_{14}^{2}}-\sqrt{\left(\mu\Delta B-\frac{1}{2}\mu(\Delta_{12}+\Delta_{34})\right)^{2}+\frac{1}{4}J_{23}^{2}}\\
-1 & \ket{\uparrow\downarrow\downarrow\downarrow} & -2\mu B_{0}+\frac{1}{2}\mu(\Delta_{12}-\Delta_{34})-\frac{1}{2}J_{14}+\sqrt{\left(\mu\Delta B+\frac{1}{2}\mu(\Delta_{12}+\Delta_{34})\right)^{2}+\frac{1}{4}J_{14}^{2}}\\
-1 & \ket{\downarrow\uparrow\downarrow\downarrow} & -2\mu B_{0}-\frac{1}{2}\mu(\Delta_{12}-\Delta_{34})-\frac{1}{2}\, J_{23}+\sqrt{\left(\mu\Delta B-\frac{1}{2}\mu(\Delta_{12}+\Delta_{34})\right)^{2}+\frac{1}{4}J_{23}^{2}}\\
-1 & \ket{\downarrow\downarrow\uparrow\downarrow} & -2\mu B_{0}-\frac{1}{2}\mu(\Delta_{12}-\Delta_{34})-\frac{1}{2}\, J_{23}-\sqrt{\left(\mu\Delta B-\frac{1}{2}\mu(\Delta_{12}+\Delta_{34})\right)^{2}+\frac{1}{4}J_{23}^{2}}\\
-1 & \ket{\downarrow\downarrow\downarrow\uparrow} & -2\mu B_{0}+\frac{1}{2}\mu(\Delta_{12}-\Delta_{34})-\frac{1}{2}J_{14}-\sqrt{\left(\mu\Delta B+\frac{1}{2}\mu(\Delta_{12}+\Delta_{34})\right)^{2}+\frac{1}{4}J_{14}^{2}}\\
-2 & \ket{\downarrow\downarrow\downarrow\downarrow} & -4\mu B_{0}
\end{array}
\]

\subsection{Logical subspace}

We can restrict our attention to the logical subspace of the two-qubit
system, as shown below; which informs us how the energy of the logical
states will change given perfectly adiabatic operation.

{\footnotesize{}
\begin{eqnarray*}
\textbf{Logical} & \textbf{Eigenstate} & \textbf{Energy}\\
\ket{11} & \ket{\uparrow\downarrow\uparrow\downarrow} & -\frac{1}{2}(J_{14}+J_{23})+\sqrt{\left(\mu\Delta B+\frac{1}{2}\mu(\Delta_{12}+\Delta_{34})\right)^{2}+\frac{1}{4}J_{14}^{2}}-\sqrt{\left(\mu\Delta B-\frac{1}{2}\mu(\Delta_{12}+\Delta_{34})\right)^{2}+\frac{1}{4}J_{23}^{2}}\\
\ket{10} & \ket{\uparrow\downarrow\downarrow\uparrow} & \mu\Delta_{12}-\mu\Delta_{34}\\
\ket{01} & \ket{\downarrow\uparrow\uparrow\downarrow} & -\mu\Delta_{12}+\mu\Delta_{34}\\
\ket{00} & \ket{\downarrow\uparrow\downarrow\uparrow} & -\frac{1}{2}(J_{14}+J_{23})-\sqrt{\left(\mu\Delta B+\frac{1}{2}\mu(\Delta_{12}+\Delta_{34})\right)^{2}+\frac{1}{4}J_{14}^{2}}+\sqrt{\left(\mu\Delta B-\frac{1}{2}\mu(\Delta_{12}+\Delta_{34})\right)^{2}+\frac{1}{4}J_{23}^{2}}
\end{eqnarray*}
}{\footnotesize \par}

Notice that the eigenvalue spectrum of these states can be reproduced
by an effective Ising model on the subspace given by:
\[
H_{\textrm{eff}}=(\mu\Delta_{12}+\bar{B})\tilde{\sigma}_{z}^{1}+(\mu\Delta_{34}+\bar{B})\tilde{\sigma}_{z}^{2}-\frac{1}{4}(J_{14}+J_{23})\left(\tilde{\sigma}_{z}^{1}\tilde{\sigma}_{z}^{2}+II\right),
\]
where $\tilde{\sigma}_{z}^{n}$ are the logical Pauli Z operators
on the logical subspace (as defined in the main text), and $\bar{B}$
is an effective magnetic field gradient between the qubits and is
given by: 
\[
\bar{B}=\frac{1}{2}\left[-\mu\Delta_{12}-\mu\Delta_{34}+\sqrt{\left(\mu\Delta B+\frac{1}{2}\mu(\Delta_{12}+\Delta_{34})\right)^{2}+\frac{1}{4}J_{14}^{2}}-\sqrt{\left(\mu\Delta B-\frac{1}{2}\mu(\Delta_{12}+\Delta_{34})\right)^{2}+\frac{1}{4}J_{23}^{2}}\right].
\]

It is worth noting that in the event that $J_{14}=J_{23}$ and in
the desired limit that $\Delta_{12},\Delta_{34}\ll\Delta B$, $\bar{B}$
reduces to $-\frac{1}{2}\mu(\Delta_{12}+\Delta_{34})$; and so:
\[
H_{\mathrm{eff}}=\frac{1}{2}\mu(\Delta_{12}-\Delta_{34})\tilde{\sigma}_{z}^{1}-\frac{1}{2}\mu(\Delta_{12}-\Delta_{34})\tilde{\sigma}_{z}^{2}-\frac{1}{4}(J_{14}+J_{23})\left(\tilde{\sigma}_{z}^{1}\tilde{\sigma}_{z}^{2}+II\right).
\]
In this limit, correcting single qubit operations amounts to keeping
track of precession due to static magnetic field gradients.

As an aside, things are not quite so simple when charge noise is added.
High frequency components of the charge noise will add uncorrectable
noise to the single qubit gates. Fortunately, since $\bar{B}\sim J^{2}$
when $J\ll\Delta B$, the errors are likely to be small. The simulations
in the main text include the high frequency single qubit errors (but
not pseudo-static noise which can be corrected); and so the reported
two-qubit gate fidelities already include the penalty for these errors.

\section{Adiabatic perturbation theory}

One of the primary sources of error for the gate described in our
paper is non-adiabatic leakage transitions that occur duing the operation
of the gate. In our paper, we present theoretical estimates for an
upper bound on leakage. This was possible because the system can be
broken down reasonably trivially into a set of two level systems,
as described in the previous section. The analytical bounds were derived
using the perturbation theory results of De Grandi and Polkovnikov
\cite{DeGrandi2010}.

For a given adiabatic pulse, the amplitude of the state $\ket{n}$
with energy $E_{n}$ at time $t_{f}$ after starting in the ground
state $\ket{0}$ with energy $E_{0}$ at time $t_{i}$ is given in
equation 19 of \cite{DeGrandi2010}:

\begin{eqnarray}
\alpha_{n}(t_{f}) & \approx & \Bigg[i\frac{\bra{n}\partial_{t}\ket{0}}{E_{n}(t)-E_{0}(t)}-\frac{1}{E_{n}(t)-E_{0}(t)}\frac{d}{dt}\frac{\bra{n}\partial_{t}\ket{0}}{E_{n}(t)-E_{0}(t)}\nonumber \\
 &  & +\ldots\Bigg]e^{i(\Theta_{n}(t)-\Theta_{0}(t))}\Bigg|_{t_{i}}^{t_{f}}\label{eq:leakage amplitude}
\end{eqnarray}

with 
\begin{eqnarray*}
\bra{n}\partial_{t}\ket{m} & = & -\frac{\bra{n}\partial_{t}\mathcal{H}\ket{m}}{E_{n}(t)-E_{m}(t)},\\
\Theta_{k}(t) & = & \int_{t_{i}}^{t_{f}}E_{k}(\tau)d\tau,
\end{eqnarray*}
and where the sequence in $\alpha_{n}(t_{f})$ is an infinite expansion
of integration by parts.

Once $\alpha_{n}$ has been computed, leakage from the ground state
is then given by: $\mathcal{L}_{0}=\sum_{n\ne0}\left|\alpha_{n}\right|^{2}$,
which is the probability of detecting a state other than the ground
state. The first non-zero contribution to $\alpha_{n}$ (which will
also be the dominant contribution in generic cases) will come from
the term that has the lowest order differential operator that when
acting on the time-dependent component of the Hamiltonian at $t_{i}$
and/or $t_{f}$ yields a non-zero value. Leakage then scales as the
square of this term. Due to the symmetry of our chosen pulses, the
constraint that each pulse must have equal area for any given gate
time, and the structure of our logical subspace, we find that the
maximum leakage error for an adiabatic pulse with first non-zero derivative
at order $q$ scales like $\left(J_{\textrm{avg}}/\Delta B\right)^{2(q+1)}$.
We derive these for the profiles used in our paper in the following
sections.

\subsection{General form of leakage probability calculations\label{sub:General-form-of}}

To simplify the derivation of leakage probabilities for each of these
pulses in the following sections, we present here a general form of
the solution. We assume that only time dependent parameter in the
model is $J_{ij}(t)$; and disregard any form of noise. Due to the
symmetry of our physical model, each two-level system has a leakage
rate determined only by the combined profile $(J_{14}+J_{23})(t)$.
We therefore write $J=J_{14}+J_{23}$ in all of these derivations.
Since $J(t)$ is the only time dependent parameter, and it acts in
each two-level system as seen in section \ref{sub:Singlet-Triplet-Qubit-System};
for all $n\ne m$, $\bra{n}\partial_{t}\ket{m}=\frac{1}{2}\frac{d}{dt}J(t)/\mu\Delta B$.

We considered in the main text adiabatic pulses with discontinuities
at differential order no greater than three; so we here expand equation
\ref{eq:leakage amplitude} to third order in derivatives of $J$.
The energy differences $\nabla_{nm}(t)=E_{n}(t)-E_{m}(t)$ will all
be approximately equal to $\mu\Delta B$; and for simplicity we drop
time dependence, since $J(t)\ll\mu\Delta B$ for any reasonable gate
operation and thus the energy eigenvalues computed in the previous
section will not vary greatly during the course of a gate. The leakage
probabalities are then given by:

\begin{eqnarray*}
\left|\alpha_{n}(t)\right|^{2} & = & \left|A_{f}+B_{f}+C_{f}-A_{i}-B_{i}-C_{i}\right|^{2}
\end{eqnarray*}

with 
\begin{eqnarray}
A_{f,i} & = & i\frac{\left\langle n|\partial_{t}|0\right\rangle }{E_{n}-E_{0}}e^{i(\theta_{n}-\theta_{0})}\bigg|_{t_{f},t_{i}}\label{eq:A}\\
B_{f,i} & = & \frac{-1}{\nabla_{n0}}\frac{d}{dt}\left(\frac{\left\langle n|\partial_{t}|0\right\rangle }{\nabla_{n0}}\right)e^{i(\theta_{n}-\theta_{0})}\bigg|_{t_{f},t_{i}}\label{eq:B}\\
C_{f,i} & = & -\frac{i}{\nabla_{n0}}\frac{d}{dt}\left(\frac{1}{\nabla_{n0}}\frac{d}{dt}\frac{\left(\bra{n}\partial_{t}\ket{0}\right)}{\nabla_{n0}}\right)e^{i(\theta_{n}-\theta_{0})}\bigg|_{t_{f},t_{i}}.\label{eq:C}
\end{eqnarray}

\subsection{Linear profile variation over J with B fixed}

In the main text we consider a linear adiabatic pulse of the form:
\[
J(t)=2J_{\textrm{avg}}\left(1-\left|\frac{2t}{\tau}-1\right|\right),
\]
with $\tau=\pi/J_{\mathrm{avg}}$.

The time derivative of $J(t)$ is:

\begin{eqnarray*}
\dot{J} & = & \begin{cases}
\frac{4}{\pi}J_{\mathrm{avg}}^{2} & t<\tau/2\\
-\frac{4}{\pi}J_{\mathrm{avg}}^{2} & t>\tau/2
\end{cases}
\end{eqnarray*}
This segmented nature of the derivative causes this pulse to have
three points of discontinuity: at the start, end and middle of the
pulse.

Since the first non-zero time derivative of $J$ is at first order;
the leading order terms in $\left|\alpha\right|^{2}$ involve the
$A_{f,i}$ terms\@. There are two segments, which under the assumptions
of constant $\nabla=\mu\Delta B$ are the same, and so we find that
leakage scales as: 
\[
|\alpha|^{2}\approx2(|A_{f}|^{2}+|A_{i}|^{2}-2Re(A_{f}A_{i}^{*})).
\]

Using equation \ref{eq:A}, $\bra{n}\partial_{t}\ket{m}=\frac{1}{2}\frac{d}{dt}J(t)/\mu\Delta B$
and $\nabla=\mu\Delta B$; we find:

\[
\left|A_{f}\right|=\left|A_{i}\right|=\frac{2}{\pi}\left(\frac{J_{\mathrm{avg}}}{\mu\Delta B}\right)^{2}
\]

Thus, the upper bound for the leakage probability (choosing the phase
in equation \ref{eq:A} to be such that $\mathrm{Re}(A_{f}A_{i}^{*})=-|A_{f}|^{2})$
is such that: 
\[
|\alpha|^{2}\lesssim\frac{32}{\pi^{2}}\left(\frac{J_{\mathrm{avg}}}{\mu\Delta B}\right)^{4}.
\]

Using the adiabatic pulse suggested as a replacement in the text $J(t)=-\frac{6\pi}{\tau^{3}}t(t-\tau)$,
one halves this upper bound.

\subsection{Sinusoidal variation of J with B fixed}

We also considered a sinusoidal pulse:
\[
J(t)=J_{\textrm{avg}}\left(1-\cos\left(\frac{2\pi t}{\tau}\right)\right),
\]
with $\tau=\pi/J_{\mathrm{avg}}$.

The time derivative of $J(t)$ is $\dot{J}(t)=2J_{\textrm{avg}}^{2}\sin\left(2J_{\textrm{avg}}t\right)$,
which is zero at initial and final times. We therefore look to the
second derivative: $\ddot{J}(t)=4J_{\textrm{avg}}^{3}\cos(2J_{\textrm{avg}}t)$.

In calculating $\left|\alpha\right|^{2}$, the leading terms are now
second order derivatives:

\[
|\alpha|^{2}\approx|B_{f}|^{2}+|B_{i}|^{2}-2Re(B_{f}B_{i}^{*}).
\]

Using equation \ref{eq:B}, $\bra{n}\partial_{t}\ket{m}=\frac{1}{2}\frac{d}{dt}J(t)/\mu\Delta B$
and $\nabla=\mu\Delta B$; we find:

\begin{eqnarray*}
|B_{f}|=|B_{i}| & = & 2\left(\frac{J_{\mathrm{avg}}}{\mu\Delta B}\right)^{3}.
\end{eqnarray*}

Thus, the upper bound for the leakage probability (choosing the phase
in equation \ref{eq:B} to be such that $\mathrm{Re}(B_{f}B_{i}^{*})=-|B_{f}|^{2})$
is such that: 
\[
|\alpha|^{2}\lesssim16\left(\frac{J_{\mathrm{avg}}}{\mu\Delta B}\right)^{6}.
\]

\subsection{XSinusoidal variation of J with B fixed}

Continuing in the trend of increasing the differential order at which
the pulse is non-zero, we also considered the so-called ``xsinusoidal''
pulse:
\[
J(t)=J_{\textrm{avg}}\frac{6\pi^{2}}{(\pi^{2}+3)}\frac{t(\tau-t)}{\tau^{2}}\left(1-\cos\left(\frac{2\pi t}{\tau}\right)\right),
\]
with $\tau=\pi/J_{\mathrm{avg}}$.

The first and second time derivatives of $J(t)$ are zero, by construction.
The third derivative, evaluated at $t=t_{i}$ or $t=t_{f}$ gives:

\[
\left|\dddot{J}\right|=72\frac{\pi}{\pi^{2}+3}J_{\textrm{avg}}^{4}.
\]

In calculating $\left|\alpha\right|^{2}$, the leading terms are now
third order derivatives:
\[
|\alpha|^{2}\approx|C_{f}|^{2}+|C_{i}|^{2}-2Re(C_{f}C_{i}^{*}).
\]

Using equation \ref{eq:C}, $\bra{n}\partial_{t}\ket{m}=\frac{1}{2}\frac{d}{dt}J(t)/\mu\Delta B$
and $\nabla=\mu\Delta B$; we find:

\begin{eqnarray*}
|C_{f}|=|C_{i}| & = & 36\frac{\pi}{\pi^{2}+3}\left(\frac{J_{\mathrm{avg}}}{\mu\Delta B}\right)^{4}.
\end{eqnarray*}

Thus, the upper bound for the leakage probability (choosing the phase
in equation \ref{eq:B} to be such that $\mathrm{Re}(C_{f}C_{i}^{*})=-|C_{f}|^{2})$
is such that: 
\[
|\alpha|^{2}\lesssim\frac{12^{4}\pi^{2}}{4(\pi^{2}+3)^{2}}\left(\frac{J_{\mathrm{avg}}}{\mu\Delta B}\right)^{8}\approx308.91\left(\frac{J_{\mathrm{avg}}}{\mu\Delta B}\right)^{8}.
\]

\section{Maximising entanglement fidelity over all single qubit z-rotations}

In the main text we mentioned that there were some subtleties regarding
how we contructed $\bar{U}$ such that we maximised the entanglement
fidelity of our gate over all single qubit z-rotations; in particular,
during the linear transformation that we perform to generate global,
single and two qubit phases, there are phase ambiguities due to sum
and differences of the extracted phases living in a larger domain.

Recall that the ansatz for our ideal unitary is $\bar{U}=e^{i\phi_{II}}e^{i\phi_{ZI}\tilde{\sigma}_{z}^{1}}e^{i\phi_{IZ}\tilde{\sigma}_{z}^{2}}e^{i\phi_{ZZ}\tilde{\sigma}_{z}^{1}\tilde{\sigma}_{z}^{2}}$.
The linear transformation which converts the phase measured in the
spin basis to the logical operator phase is given by: 
\begin{equation}
\left(\begin{array}{c}
\phi_{II}\\
\phi_{IZ}\\
\phi_{ZI}\\
\phi_{ZZ}
\end{array}\right)=\left(\begin{array}{cccc}
1 & 1 & 1 & 1\\
1 & -1 & 1 & -1\\
1 & 1 & -1 & -1\\
1 & -1 & -1 & 1
\end{array}\right)\left(\begin{array}{c}
\Phi_{00}\\
\Phi_{01}\\
\Phi_{10}\\
\Phi_{11}
\end{array}\right).\label{eq:phase_transformation}
\end{equation}
At this stage, each of the logical phases $\phi_{xy}$ are elements
of the domain $[-4\pi,4\pi)$; whereas we only care about their value
modulo $2\pi$. If we were simply to invert this relation, we would
extract the original phases in the spin basis; but when we enforce
the two-qubit phase to be $\pi\in[-2\pi,2\pi)$, there is an ambiguity
as to which value of $\phi_{ZZ}\in[-4\pi,4\pi)$ should be selected.

In experiment this would not be a problem, because one would simply
keep track of the accumulated single qubit phases and then correct
them appropriately; but in our simulations, we did not want to have
to keep track of extra state information. To avoid this ambiguity,
we simply considered all four possible values of $\phi_{ZZ}=\pi\mod2\pi\in[-4\pi,4\pi)$:
$-3\pi$, $-\pi$, $\pi$, and $3\pi$; taking the supremum of the
associated entanglement fidelities (computed as described in the main
text).

\section{Origin of the Fidelity-Leakage relation}

In our paper we claim, without proof, the fidelity-leakage relation
shown in equation (4): 
\[
\mathcal{F}=\frac{1}{2}\left(1+\sqrt{1-2\mathcal{L}_{0}}\cos\left(2\Delta\right)-\mathcal{L}_{0}\right),
\]
where $\Delta=\phi_{ZZ}-\bar{\phi}_{ZZ}$, which characterises any
under- or over- accrual of two-qubit phase acquired during the gate
operation. We demonstrate that this is a simple corrollary of the
symmetries of our model.

Recall that there are exactly two leakage states: $\ket{\uparrow\uparrow\downarrow\downarrow}$
and $\ket{\downarrow\downarrow\uparrow\uparrow}$. Excitations to
these states occur from the two logical states: $\ket{\uparrow\downarrow\uparrow\downarrow}$
and $\ket{\downarrow\uparrow\downarrow\uparrow}$, under the action
of the inter-qubit exchange couplings associated with $J_{14}$ and
$J_{23}$. An examination of the analysis in section \ref{sec:Analytically-solving-for}
of this supplementary material shows that the leakage rates depend
on exchange coupling $J_{ij}$. Moreover, the leakage is symmetrical,
in that both exchange couplings generate leakage equally into both
leakage states and from both logical states.

We can therefore write a general ansatz for the state of the system
after some time evolution starting from the maximally entangled state
$\left|\Psi\right\rangle $:
\begin{eqnarray*}
\ket{\psi} & = & \frac{1}{2}\bigg(p\exp(i\phi_{l})\ket{\uparrow\uparrow\downarrow\downarrow}+\sqrt{1-p{}^{2}}\exp(i\phi_{11})\ket{\uparrow\downarrow\uparrow\downarrow}+\exp(i\phi_{10})\ket{\uparrow\downarrow\downarrow\uparrow}\\
 &  & +\exp(i\phi_{01})\ket{\downarrow\uparrow\uparrow\downarrow}+\sqrt{1-p^{2}}\exp(i\phi_{00})\ket{\downarrow\uparrow\downarrow\uparrow}+p\exp(i\phi_{m})\ket{\downarrow\downarrow\uparrow\uparrow}\bigg).
\end{eqnarray*}
This state has leakage given by $\mathcal{L}=p^{2}/2$.

Suppose now that we constructed an ideal state $\ket{\bar{\psi}}$
that has evolved from the same maximally entangled state such that
$\bar{U}$ has been applied, as described in the previous section.
By construction, the only component of these states which will differ
is their two-qubit phase. Taking their inner product, it can be shown
that:
\[
\inner{\bar{\psi}}{\psi}=\frac{1}{2}\left(\exp(-i\Delta)+\sqrt{1-2\mathcal{L}}\exp(i\Delta)\right).
\]
The result then follows from the definition of entanglement fidelity
given in the text.

\end{document}